\documentclass[aps,prfluids,preprint,superscriptaddress]{revtex4-2}
\usepackage{amsmath,amssymb} 
\usepackage{color}
\usepackage[pdfstartview=FitH,      
breaklinks=true,        
bookmarksopen=true,     
bookmarksnumbered=true  
]{hyperref}
\hypersetup{colorlinks=true,allcolors=blue}
\usepackage{float}
\usepackage{bm}
\usepackage{cases}
\usepackage{graphicx}
\begin{document}


\title{Generation of interfacial waves by rotating magnetic fields}


\author{Gerrit Maik Horstmann}
\email[]{g.horstmann@hzdr.de}
\affiliation{Institute of Fluid Dynamics, Helmholtz-Zentrum Dresden-Rossendorf, Bautzner Landstrasse 400, D-01328 Dresden, Germany}
\author{Yakov Nezihovski}
\affiliation{School of Mechanical Engineering, Faculty of Engineering, Tel Aviv University, Tel Aviv 6997801, Israel}
\author{Thomas Gundrum}
\affiliation{Institute of Fluid Dynamics, Helmholtz-Zentrum Dresden-Rossendorf, Bautzner Landstrasse 400, D-01328 Dresden, Germany}
\author{Alexander Gelfgat}%
\affiliation{School of Mechanical Engineering, Faculty of Engineering, Tel Aviv University, Tel Aviv 6997801, Israel}%
\date{\today}
\begin{abstract}
Interfacial waves arising in a two-phase swirling flow driven by a low-frequency rotating magnetic field (RMF) are studied. At low RMF frequencies, of the order of 1-$10\,{\rm Hz}$, the oscillatory part of the induced Lorenz force becomes comparable to the time-averaged one, and cannot be neglected. In particular, when free surfaces or two-liquid stably stratified systems are subject to a low-frequency RMF, induced pressure variations necessarily excite free-surface/interfacial waves, which can improve mass transfer in different metallurgical processes. 
In this paper, we formulate a linear wave model and derive explicit analytical solutions predicting  RMF-driven wave patterns that closely resemble hyperbolic paraboloids. These theoretical predictions are validated against experiments based on a non-intrusive acoustic measurement technique, which measures liquid-liquid interface elevations in a two-phase KOH-GaInSn stably stratified system. A good quantitative agreement is found for non-resonant wave responses in the vicinity of the fundamental resonance frequency. The experiments reveal the additional excitation of several higher harmonics superimposing the fundamental wave oscillation, which are visible even in the linear wave regime.	 
\end{abstract}


\maketitle

\section{Introduction}
\label{sec:Introduction}
Rotating magnetic fields (RMF) are widely used in metallurgy as they allow contactless stirring and mixing of liquid melts under controllable conditions \cite{Gelfgat1995,Davidson2001}. In continuous casting, as one prominent example, electromagnetic stirring is assumed to enhance the homogeneity of the molten steel and can thereby reduce the number of casting defects \cite{Spitzer1986,Kunstreich2003}. In semiconductor industries, RMF find likewise application and are used to control heat and mass transfer in single crystal growth both in the melt and at the crystallization front \cite{Gelfgat1999a,Gelfgat2001}. Apart from the wide range of applications, RMF-driven flows are also very attractive from an academic perspective, explaining why RMF stirring has evolved to a textbook example in the field of Magnetohydrodynamics \cite{Davidson2001}. A tremendous effort has been made over the last 50 years to understand the intricate flow physics and the mechanisms of action that RMF have on electrically conducting liquids \cite{Moffatt1965,Dahlberg1972,Davidson1987,Sneyd1993,Priede1996,Witkowski1998, Grants2001,Grants2002}. RMF flows take on such interest because they involve various secondary flows alongside the dominant azimuthal swirling flow. Perhaps most importantly, the lower (and possibly upper) end of the usually cylindrical stirring vessel creates a meridional flow driven by the Ekman pumping mechanism \cite{Davidson1995}, which can be of the same scale as the primary azimuthal flow and facilitates vertical mixing. Also worth mentioning is the occurrence of recirculating flows, resulting from often inevitable axial variations in the stirring force, as well as different types of flow instabilities and the presence of Taylor-G\"{o}rtler vortices forming at higher Taylor numbers. More recent studies are mostly dedicated to the transient spin-up phase of the liquid metal flow \cite{Nikrityuk2004,Nikrityuk2005,Rabiger2010,Vogt2012,Travnikov2012}
or spin-up of a concentrated vortex forming at the metal free-surface in response to pulses of traveling magnetic fields (TMF) superimposed to the RMF \cite{Grants2008,Vogt2013,Grants2015}.

In several electromagnetic stirring applications, as in continuous casting, the stirring vessel is not closed at the top, but instead the melt forms a free surface in contact with air, which is set in motion by the RMF in tandem with the bulk liquid. The motion and composition of the free surface can have a significant effect on the metallurgical process. For example,
the transitions from steady to oscillatory flow regimes depend
sensitively on the cleanliness of the liquid metal surface \cite{Nikrityuk2004,Travnikov2012}. Also, a tornado-like vortex can be driven by a combination of rotating and traveling magnetic fields that leads to funnel-like depression \cite{Grants2008,Vogt2013} of the free surface and allows the entrapment of floating additives (unwetted particles) into the molten metal \cite{Gelfgat2005}. Such an entrainment of solid particles can sometimes be beneficial in metallurgy, e.g., for alloying, preparing specific melts or to re-melt scrap. In continuous casting, however, strong surface deformations intensifying the entrapment of 
slag or mould powder are to be avoided since impurities can lower the quality of steel products considerably  \cite{Willers2017}. For these reasons, surface displacements that accompany the swirling flow have come to the fore in several studies, but invariably only with regard to the RMF stirring force. 

In general, the RMF-induced Lorentz force consists of two parts: a mean (time-averaged) part that drives the swirling flow and an oscillatory part that is mostly neglected in the literature. However, in the range of low RMF frequencies, i.e.\ from about 1 to $10\, {\rm Hz}$\textemdash the typical frequency range applied in industrial mould stirrers\textemdash the oscillatory part cannot be disregarded as it generates gravity (or capillary) waves on the free surface or interface. Such waves have already been investigated for other magnetic field arrangements, most prominent are the studies by \cite{Galpin1992,Galpin1992a,Fautrelle2005a,Fautrelle2007,Deng2011}, who applied alternating vertical magnetic fields to free-surface liquid metal pools. The authors found that alternating magnetic fields can both excite axisymmetric standing wave modes, which are direct solutions of the forced wave problem, as well
as non-symmetric azimuthal wave modes resulting from a parametric instability. More recently, pulsed magnetic fields have been applied to excite surface waves as well \cite{Wu2020,Milgravis2023}. Such types of magnetic field-induced irrotational wave motions are known to have some metallurgically favorable properties and can, in a similar way to free surface motions driven by the mean part of the RMF Lorentz force, improve mass transfer as they increase the surface area \cite{Debray1996,Milgravis2020}. 

Despite the apparently large effects of surface motions, interfacial waves excited by RMF have not yet been investigated to our best knowledge, with the notable exception of the dissertation by \citet{Wiederhold2019}, where a rotating disk equipped with a permanent magnet was placed beneath a cylindrical container filled with the eutectic alloy GaInSn. This setup allowed strong rotating wave motions to be excited whenever resonance conditions could be established. Yet, the magnetic field, which decreases exponentially in the axial direction, is highly inhomogeneous and greatly complicates the wave physics. This is precisely the point at which we intend to embark this study and investigate interfacial waves that are excited by a vertically homogeneous RMF, which is much more accessible for theoretical modeling. For this purpose, we consider an idealized setup of an upright circular cylinder placed concentrically in an homogeneous RMF and filled by two immiscible electrically conducting liquids forming a two-layer stably stratified system. In Sec.\,\ref{sec:Model} we formulate a wave model, which comprises the irrotational oscillatory part of the Lorentz force and can account for both magnetically excited gravity-capillary free-surface and interfacial waves. Explicit analytic solutions are derived for leading-order surface elevations. In Sec.\,\ref{sec:Exp}, we present a novel magnetohydrodynamic wave experiment, in which a specific arrangement of induction coils is utilized to generate a virtually homogeneous RMF that excites rotating waves on the interface formed between stratified KOH and GaInSn liquid layers. Interface elevations are measured acoustically through an arrangement of up to ten ultrasonic sensors. These measurements are fully non-intrusive.
Finally, different types of observed non-resonant and resonant wave motions are thoroughly discussed and compared to the theoretical predictions in Sec.\,\ref{Sec:Results}.     
\newpage
\section{Theoretical model}
\label{sec:Model}
\subsection{Mathematical framework}
\begin{figure}
	\vspace*{-1.35cm}
	\centering
	\def\svgwidth{250pt}    
	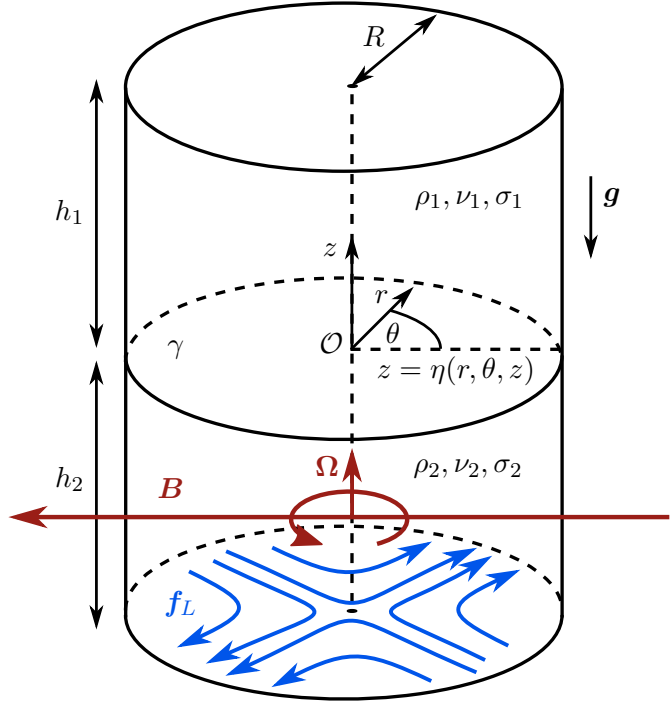
	\vspace*{-0.4cm}	
	\caption{Sketch of the theoretical setup. An upright cylindrical container of radius $R$ is permeated by an external homogeneous magnetic field $\bm{B}$ rotating horizontally around the $z$-axis with constant angular frequency $\bm{\Omega} = \Omega \bm{e}_z$.
		The container is filled with two immiscible  liquids $i = 1, 2$ of densities $\rho_i$ , kinematic viscosities $\nu_i$, electrical conductivities $\sigma_i$ and layer heights $h_i$,
		which are stably stratified due to gravity $\bm{g}$. The origin of the cylindrical coordiante system in placed in the center of the interface $z = \eta(r,\theta,z)$. The blue arrows schematically show force lines of the conservative part of the induced Lorentz force.}
	\label{fig:Setup}
\end{figure}
The theoretical framework to be treated in this study is illustrated in Fig.\,\ref{fig:Setup}. We define an ideal circular cylinder of radius $R$, which shall embody two immiscible liquid phases (subscripts $i=1,2$) specified by different densities $\rho_1$, $\rho_2$, kinematic viscosities $\nu_1 , \nu_2$ and electrical conductivities $\sigma_1$, $\sigma_2$, where $\rho_1 < \rho_2$ must be fulfilled to ensure a stable vertical stratification. We align the axis of symmetry with the $z$-axis of a cylindrical coordinate system ($r, \theta, z$) with unit vectors ($\bm{e}_r , \bm{e}_{\theta}, \bm{e}_z$). At equilibrium, the two phases occupy the fluid domains 
\begin{align}
&\mathcal{V}_1 : (r,\theta,z)\in [0,R]\times [0,2\pi)\times [0,h_1], \\
&\mathcal{V}_2 : (r,\theta,z)\in [0,R]\times [0,2\pi)\times [0,h_2],
\end{align}
where $h_1$ and $h_2$ are the heights of the two layers. The interface between both phases is
placed at $z = \eta(r, \theta, t)$, the coordinate origin $\mathcal{O}$ is defined in the center of the equilibrium interface. We incorporate interfacial tension $\gamma$ for the liquid-liquid interface but neglect capillary effects in the contact line region, i.e., the interface is assumed to slide freely along the cylinder wall while maintaining a static contact angle of $90^{\circ}$ (no meniscus). Further, two independent volume forces are taken into account to act on the system. First, we need to consider gravity pointing in the negative $z$-direction ($\bm{g} = -g \bm{e}_z$), which manifests itself as a restoring force. Second, the entire cylinder is horizontally permeated by an external homogeneously rotating magnetic field $\bm{B}$ given as
\begin{align}
	\bm{B} (\theta,t) = 
	\left(
	\begin{array}{r}
		B_0 \sin(\Omega t - \theta)\bm{e}_{r}\\
		-B_0 \cos(\Omega t - \theta)\bm{e}_{\theta}\\
	\end{array}
	\right), \label{eq:RMF}
\end{align}
where $\Omega$ is the field's angular frequency. This RMF first induces closing electrical currents in the conductive layers via Ohm's law, which then interact again with the RMF and create a rotating Lorentz force. In general, the Lorentz force can be decomposed into a mean part, which drives an axisymmetric swirling flow widely established in stirring applications, and an oscillatory part mostly neglected in the literature. The oscillatory part itself can again be split into rotational and irrotational parts. The corresponding mathematical expressions are given below. In this work, we are going to concern exactly the latter part (visualized by the blue force lines in Fig. \ref{fig:Setup}), that can excite irrotational wave motion on the interface in exactly the same manner as classic conservative forces as, e.g., fictitious inertia forces appearing in shaken containers, cause sloshing waves. 

This formulation is characterized by thirteen physical variables and four physical
dimensions. Following the Buckingham $\Pi$ theorem, the system can be uniquely described by nine independent dimensionless quantities. We define the following set of dimensionless numbers for our two-layer ($i=1,2$) analysis:
\begin{gather}
	Fr = \frac{|\sigma_2 - \sigma_1| \Omega B_0^2 R}{(\rho_2 - \rho_1)g}, \,N_i = \frac{\sigma_i B_0^2}{\Omega (\rho_2 - \rho_1)}, \,Re_i = \frac{\Omega R^2}{\nu_i}, \, H_i = \frac{h_i}{R}, \nonumber \\
	 Bo = \frac{(\rho_2 - \rho_1)gR^2}{\gamma}, \,A = \frac{\rho_2 - \rho_1}{\rho_1 + \rho_2}. \label{eq:Numbers}
\end{gather}
The magnetic Froude number $Fr$ describes the ratio of the Lorentz force per unit mass to the restoring gravity force acting on the interface. The numbers $N_i$ are the phase-dependent magnetic interaction parameters (also called Stuart number) and are a measure for the impact Lorentz forces can have on the hydrodynamic interface. The Reynolds numbers $Re_i$ are also phase dependent and are here expressed in a way that they weight the cylinder radius with the characteristic Stokes boundary layer thicknesses $\delta_i \sim \sqrt{\nu_i / \Omega}$. The importance of gravitational forces compared with interfacial tension forces to the wave motion is quantified by the Bond number $Bo$. Finally, $H_i$ and $A$ are the layer aspect ratios and the Atwood number, which describes the transition from one-layer free-surface waves ($A=1$) to two-layer interfacial waves ($A \ll 1$).
\subsection{Treatment of the Lorentz force}
We evaluate the induced Lorentz force in the framework of the so called inductionless (also magnetostatic) low-frequency approximation, demanding that both the magnetic Reynolds number $Re_m = u R \mu_0 \sigma$ and the shielding (also skin depth) parameter $\Delta = \Omega R^2 \mu_0 \sigma$ are small $Re_m < \Delta \ll 1$. Small magnetic Reynolds numbers allow us to ignore advection of the magnetic field so that the liquids can be treated as stationary solid conductors. Small shielding parameters ensure that $\bm{B}$ can completely and sufficiently fast permeate into the liquids, i.e., the skin depth $\delta_s =\sqrt{2/\mu_0 \sigma \Omega}$ must be large relative to $R$. This approximation allows to uniquely describe the electric field $\bm{E}$ through the gradient of a scalar electrical potential $\bm{E} = - \bm{\nabla}\varphi$ and the induced current $\bm{j}$ is calculated through Ohm's law in both liquids
\begin{align}
	\bm{j}_i = \sigma_i (-\bm{\nabla}\varphi_i + \bm{u}_i\times \bm{B}).
\end{align}
For calculating the Lorentz component $\bm{u}\times \bm{B}$, we can make use of the fact that a magnetic field rotating in a resting cylinder is equivalent to an oppositely rotating cylinder within a fixed magnetic field. As a result of the magnetic stirring, however, a angular flow will develop in the liquid, which reduces the rotational difference between the field $\sim r\Omega$ and the fluid $v$. If the motion is approximated as being independent of the height $z$, the difference in angular velocity $r\Omega - v$ is the key driving parameter. Most generally, when considering finite cylinders, rotating liquids are known to induces secondary flows so that all three velocity components $(u,v,w)$ contribute to the Lorentz force. The induced currents then yield \citep{Tagawa2019}:
\begin{align}
\bm{j}_{r,i} (r,\theta,z,t) = 
\left(
\begin{array}{l}
	\sigma_i \left(-\frac{\partial \varphi_i}{\partial r} + w_i B_0 \cos(\Omega t - \theta )\right) \bm{e}_{r}\\
	\sigma_i \left(-\frac{1}{r}\frac{\partial \varphi_i}{\partial \theta} + w_i B_0 \sin(\Omega t - \theta )\right)\bm{e}_{\theta}\\
	\sigma_i \left(-\frac{\partial \varphi_i}{\partial z} + \left(r\Omega - v_i \right) B_0 \sin(\Omega t - \theta ) - u_i B_0 \cos(\Omega t - \theta ) \right)\bm{e}_z
\end{array}
\right). \label{eq:Current}	
\end{align}   
The electrical potentials must be formally calculated by solving the Poisson equation. Yet, in some cases, where the influence of the lower and upper boundaries can be neglected, no significant electric fields are generated and the potential contributions can be omitted. In our case of low-frequency excitation, we can assume that secondary flows $u,w$ caused by bulk electromagnetic stirring are negligibly small as compared to wave-induced motions ($u,w \ll r \Omega$). This is not necessarily true for the primary azimuthal component $v$, which is smaller than the angular velocity of the field $v < r\Omega$ but finite. The resulting flow will therefore be a superposition of swirling flow and the rotating wave motion to be calculated, meaning that the wave will co-rotate with the stirred liquid metal and additionally propagate within this frame of reference with its wave velocity. The impact of $v$, however, on the irrotational part of the Lorentz force driving the wave motion is small and therefore also be neglected in the following. Given these assumptions, a simplified form of the electromagnetic force can be stated as \cite{Tagawa2019}: 
\begin{align}
	\bm{f}_i (r,\theta,t) = 
	\frac{\sigma_i B_0^2}{2}\left(
	\begin{array}{l}
		r\Omega\sin(2\Omega t - 2\theta)\bm{e}_r \\
		r\Omega(1-\cos(2\Omega t - 2\theta))\bm{e}_{\theta}
	\end{array}
	\right). \label{eq:Lorentz}	
\end{align}
This force field can now be divided into a periodic oscillatory and a non-periodic averaged part. The averaged Lorentz force is purely azimuthal at first approximation   
\begin{align}
 \langle f_i (r) \rangle = \frac{\sigma B_0^2}{2}r\Omega \bm{e}_{\theta} \label{Eq:FMean}
\end{align}
and is the key component underlying magnetic stirring used in several industrial applications. The oscillatory part, in contrast, is widely disregarded in the literature since most studies consider RMF frequencies of 50 or $60\, {\rm Hz}$, where the oscillatory component has no significant impact. In this study, we are examining small RMF frequencies below $10\,{\rm Hz}$. In this range, both force components are of comparable magnitude, but still the oscillatory part cannot drive any flow in one-phase fluids and only produces magnetic pressure \citep{Dahlberg1972,Davidson1987}. If, however, a free surface or interface is present in the system, it is exactly the irrotational part causing wave motion because the magnetic pressure is balanced by the hydrostatic pressure resulting from free surface/interface elevations. The oscillatory part is irrotational and can be uniquely expressed as a gradient of a scalar potential $\phi_{L,i}$ as $\bm{f}_i = \bm{\nabla}\phi_{L,i}$, where
\begin{align}
	\phi_{L,i} = \frac{\sigma_i B_0^2}{4}\Omega r^2 \sin(2\Omega t - 2\theta). \label{eq:PotL}
\end{align}
This Lorentz force potential resembles the centrifugal potential commonly used to describe sloshing waves in orbitally shaken containers \citep{Horstmann2020}, with the fine difference that it involves a $2\theta$ (instead of $1\theta$) azimuthal periodicity, exciting waves with two nodal diameters (two crest-trough pairs along the circumference) instead of one nodal diameter in the leading order, as we are going to show in the following. 

\subsection{Statement of the hydrodynamic boundary value problem}
We formulate the wave problem within the framework of potential theory, demanding that the flow fields are irrotational ($\bm{\nabla}\times \bm{u}_i = 0$). This approximation is justified for not too small Reynolds numbers $Re_i \gtrsim 100$, ensuring that the boundary layers thickness $\delta_i \sim \sqrt{\nu_i /\Omega}$ is orders of magnitude smaller than the characteristic
length scales $\sim R$ of the wave motion, such that the rotational part of the flow is confined  in the close vicinity of the tank walls and the interface. Under the assumption of irrotationality, all flows fields can, exactly as the conservative part of the Lorentz force, be uniquely expressed through the gradients of  scalar flow potentials $\bm{u}_i = \bm{\nabla}\phi_i$. This way, the $2\times 3$ velocity components $(u_{r,1} , u_{\theta,1}, u_{z,1})$ and $(u_{r,2} , u_{\theta,2}, u_{z,2})$ have been reduced by four degrees of freedom to two flow potentials $\phi_1$ and $\phi_2$, whereby the problem is mathematically greatly simplified. As the last simplification, we only seek for first-order solutions requiring that the wave amplitude $\eta_0$ is small as compared to the lateral dimensions $\eta_0 \ll R$. Then, the
linear wave problem can be stated by the following complete set of linear equations:
\begin{subnumcases}{\label{Eq:Set}}
	\frac{\partial \phi_i}{\partial t}  - \frac{\sigma_i B_0^2}{4\rho_i}\Omega r^2 \sin(2\Omega t - 2\theta) + \frac{p_i}{\rho_i} + g z = c_i (t), & ({\rm Flow \ fields}) \label{Eq:a} \\
	\Delta \phi_i = \frac{\partial^2 \phi_i}{\partial r^2} + \frac{1}{r}\frac{\partial \phi_i}{\partial r} + \frac{1}{r^2}\frac{\partial^2 \phi_i}{\partial \theta^2} + \frac{\partial^2 \phi_i}{\partial z^2} = 0, & ({\rm Flow \ fields}) \label{Eq:b}\\
	\frac{\partial \phi_1}{\partial z} = 0|_{z = h_1}, & ({\rm Top \ wall})  \label{Eq:c} \\
	\frac{\partial \phi_2}{\partial z} = 0|_{z = - h_2}, & ({\rm Bottom \ wall}) \label{Eq:d}\\
	\frac{\partial \phi_1}{\partial r} = \frac{\partial \phi_2}{\partial r} = 0|_{r = R}, & ({\rm Side \ wall}) \label{Eq:e}\\
	\frac{\partial \phi_1}{\partial z} = \frac{\partial \phi_2}{\partial z} = \frac{\partial \eta}{\partial t} |_{z=0}, & ({\rm Interface}) \label{Eq:f} \\
	\gamma \Delta_H \eta = \gamma \left(\frac{\partial^2 \phi_i}{\partial r^2} + \frac{1}{r}\frac{\partial \phi_i}{\partial r} + \frac{1}{r^2}\frac{\partial^2 \phi_i}{\partial \theta^2}\right)\eta =  p_1 - p_2|_{z=0}. & ({\rm Interface}) \label{Eq:g} 
\end{subnumcases}
The first two equations (\ref{Eq:a}) and (\ref{Eq:b}) are the instationary Bernoulli equation and
the Laplace equation ensuring energy and mass conservation in both layers. Equations (\ref{Eq:c}), (\ref{Eq:d})
and (\ref{Eq:e}) comprise the kinematic no-outflow boundary conditions at the cylinder
walls. Equation (\ref{Eq:f}) is an additional kinematic boundary condition achieving the
preservation of the interface. Finally, the formulation is closed by the linearized Young-Laplace
equation (\ref{Eq:g}) relating the pressure discontinuity at the interface to the capillary
pressure. In the Bernoulli equations, $c_1 (t)$ and $c_2 (t)$ are arbitrary integration constants, which can be incorporated into the force potentials $\tilde{\phi}_{L,i} = \phi_{L,i} + c_i (t)$ and do not need to be considered further. A detailed derivation of the system (\ref{Eq:Set}) (without Lorentz forcing) can be found in \cite[Chapter~2.2.1]{Horstmann2021a}.
\subsection{Modal equations and solutions}
A convenient way to approach this problem is to first solve the Laplace equations (\ref{Eq:b})
together with the kinematic conditions (\ref{Eq:c}) - (\ref{Eq:f}).
The solutions can be expanded as a series of harmonic, hyperbolic and Bessel functions in the following way (see \citep[Chapter~A.1]{Horstmann2021a}):
\begin{align}
	&\phi_{1}(r,\theta,z,t) = - \sum_{m=0}^{\infty}\sum_{n=1}^{\infty}\Phi_{mn}(\theta, t)\frac{\cosh \left(\frac{\epsilon_{mn}}{R}(z-h_1)\right)}{\sinh \left(\frac{\epsilon_{mn}}{R} h_1 \right)}J_m \left(\frac{\epsilon_{mn}r}{R}\right), \label{eq:Pot1} \\
	&\phi_{2}(r,\theta,z,t) = \sum_{m=0}^{\infty}\sum_{n=1}^{\infty}\Phi_{mn}(\theta, t)\frac{\cosh \left(\frac{\epsilon_{mn}}{R}(z+ h_2)\right)}{\sinh \left(\frac{\epsilon_{mn}}{R} h_2 \right)}J_m \left(\frac{\epsilon_{mn}r}{R}\right),  \label{eq:Pot2} \\
	&{\rm with} \ \ \Phi_{mn}(\theta,t) = \alpha_{mn}(t)\cos(m\theta) + \beta_{mn}(t)\sin(m\theta), \nonumber
\end{align}
where $J_m$ are the $m$-th order Bessel functions of the first kind and $\epsilon_{mn}$ denote the mode-dependent (radial)
wave numbers, restricted to the $n$ roots of the first derivative of the $m$-th order Bessel
function $J'_m(\epsilon_{mn}) = 0$ in order to satisfy the no-outflow condition at the sidewalls.
The integers $n \in \mathbb{N}_1$ and $m \in \mathbb{N}_0$ indicate the number of antinodal cycles (crest-trough pairs along the diameter) and antinodal diameters (crest-trough pairs within one antinodal cycle along the circumference). Finally $\alpha_{mn}(t)$ and $\beta_{mn}(t)$ are the modal functions entailing the time dependence and are yet to be determined. The change of the flow potentials in time is described by equations (\ref{Eq:a}), allowing us to calculate the induced pressures $p_1$ and $p_2$, which are balanced by the capillary pressure at the interface (\ref{Eq:g}). Eliminating the pressure difference in (\ref{Eq:g}) yields
\begin{align}
\gamma \Delta_H \eta = \rho_2 \frac{\partial \phi_2 }{\partial t} - \rho_1 \frac{\partial \phi_1  }{\partial t} + \frac{(\sigma_1 - \sigma_2) B_0^2 \Omega}{4}r^2 \sin(2\Omega t - 2\theta) + (\rho_2 - \rho_1)g\eta|_{z=0} \label{eq:Dyn1}
\end{align}  
Next, we can eliminate the interface elevation $\eta$ by differentiating equation (\ref{eq:Dyn1}) with respect to time and using equation (\ref{Eq:f}), giving
\begin{align}
 \rho_2 \frac{\partial^2 \phi_2 }{\partial t^2} - \rho_1 \frac{\partial^2 \phi_1  }{\partial t^2} + \frac{(\sigma_1 - \sigma_2) B_0^2 \Omega^2}{2} r^2 \cos(2\Omega t - 2\theta) + (\rho_2 - \rho_1)g\frac{\partial \phi_1}{\partial z} + \gamma \frac{\partial^3 \phi_1}{\partial z^3
 } = 0|_{z=0} \label{eq:Dyn2}	
\end{align}
The horizontal Laplacian was removed through the continuity equation $\partial_z^2 \phi_i = - \Delta_H \phi_i$. Now, the only remaining problem is that the Lorentz force potential does not conform to the function basis of the flow potentials (\ref{eq:Pot1}) and (\ref{eq:Pot2}). The force potential must therefore be projected onto the Fourier-Bessel eigenbasis of the flow potentials, here easily achieved by expressing the radial component $r^2$ in terms of a Bessel series as
\begin{align}
	r^2 = \sum_{n=1}^{\infty}\frac{4R^2 J_2\left(\frac{\epsilon_{2n}}{R}r\right)}{(\epsilon_{2n}^2 - 4)J_2 (\epsilon_{2n})}. \label{eq:Bessel}
\end{align}
Inserting (\ref{eq:Bessel}) and the potential solutions (\ref{eq:Pot1}) and (\ref{eq:Pot2}) into equation (\ref{eq:Dyn2}) allows to derive conditions for the modal coefficients, we find
\begin{eqnarray}
	&&\sum_{m=0}^{\infty}\sum_{n=1}^{\infty}\rho_1 \coth\left(\frac{\epsilon_{mn}}{R} h_1\right) \left[\ddot{\alpha}_{mn}(t)\cos(m\theta) + \ddot{\beta}_{mn}(t)\sin(m\theta)\right]J_m \left(\frac{\epsilon_{mn}r}{R}\right) \nonumber \\
	&&\sum_{m=0}^{\infty}\sum_{n=1}^{\infty}\rho_2 \coth\left(\frac{\epsilon_{mn}}{R} h_2\right)\left[\ddot{\alpha}_{mn}(t)\cos(m\theta) + \ddot{\beta}_{mn}(t)\sin(m\theta)\right]J_m \left(\frac{\epsilon_{mn}r}{R}\right) \nonumber \\
	&&\sum_{m=0}^{\infty}\sum_{n=1}^{\infty}(\rho_2 - \rho_1)g\frac{\epsilon_{mn}}{R}\left[\alpha_{mn}(t)\cos(m\theta) + \beta_{mn}(t)\sin(m\theta)\right]J_m \left(\frac{\epsilon_{mn}r}{R}\right) \nonumber \\
	&&\sum_{m=0}^{\infty}\sum_{n=1}^{\infty}\gamma\left(\frac{\epsilon_{mn}}{R}\right)^3\left[\alpha_{mn}(t)\cos(m\theta) + \beta_{mn}(t)\sin(m\theta)\right]J_m \left(\frac{\epsilon_{mn}r}{R}\right) \nonumber \\
	&&= -\sum_{n=1}^{\infty}\frac{2(\sigma_1 - \sigma_2) B_0^2 R^2\Omega^2 }{(\epsilon_{2n}^2 - 4)J_2 (\epsilon_{2n})}\left[\cos(2\Omega t)\cos(2\theta) + \sin(2\Omega t)\sin(2\theta)\right]J_2\left(\frac{\epsilon_{2n}}{R}r\right). \label{eq:Modal}
\end{eqnarray}
This equation can only be fulfilled for
\begin{align}
\alpha_{mn}(t) = \beta_{mn}(t) = 0 \  {\rm if} \ m\neq 2.	
\end{align}
For the modes $m=2$, equation (\ref{eq:Modal}) can be rearranged into the following form
\begin{align}
&\sum_{n=1}^{\infty}\Bigg[\left(\rho_1 \coth\left(\frac{\epsilon_{mn}}{R} h_1\right) + \rho_2\coth\left(\frac{\epsilon_{mn}}{R} h_2\right)\right)\ddot{\alpha}_{mn}(t) + (\rho_2 - \rho_1)g\frac{\epsilon_{mn}}{R}\alpha_{mn}(t) \Bigg.
\nonumber \\
+ &\Bigg. \gamma\left(\frac{\epsilon_{mn}}{R}\right)^3 \alpha_{mn}(t) + \frac{2(\sigma_1 - \sigma_2) B_0^2 R^2\Omega^2 }{(\epsilon_{2n}^2 - 4)J_2 (\epsilon_{2n})}\cos(2\Omega t) \Bigg]\cos(2\theta)J_2\left(\frac{\epsilon_{2n}}{R}r\right) \nonumber \\ 
&\sum_{n=1}^{\infty}\Bigg[\left(\rho_1 \coth\left(\frac{\epsilon_{mn}}{R} h_1\right)+ \rho_2\coth\left(\frac{\epsilon_{mn}}{R} h_2\right)\right)\ddot{\beta}_{mn}(t)  + (\rho_2 - \rho_1)g\frac{\epsilon_{mn}}{R}\beta_{mn}(t) \Bigg.
\nonumber \\
+ &\Bigg. \gamma\left(\frac{\epsilon_{mn}}{R}\right)^3 \beta_{mn}(t)+ \frac{2(\sigma_1 - \sigma_2) B_0^2 R^2\Omega^2 }{(\epsilon_{2n}^2 - 4)J_2 (\epsilon_{2n})}\sin(2\Omega t) \Bigg]\sin(2\theta)J_2\left(\frac{\epsilon_{2n}}{R}r\right) = 0.
\end{align}
The infinite sums can only yield zero if all individual summands disappear. Therefore,
each coefficient $\alpha_{2n}(t)$ and $\beta_{2n}(t)$ must satisfy the following set of modal equations:
\begin{align}
	&\ddot{\alpha}_{2n}(t) + \omega_{2n}^2 \alpha_{2n}(t) + F_{2n}\cos(2\Omega t) = 0, \label{Eq:ModalA}\\
	&\ddot{\beta}_{2n}(t) + \omega_{2n}^2 \beta_{2n}(t) + F_{2n}\sin(2\Omega t)=0, \label{Eq:ModalB}
\end{align}
where
\begin{equation}
	\omega_{2n}^{2} = \frac{(\rho_2 - \rho_1)g\frac{\epsilon_{2n}}{R} + \gamma\left(\frac{\epsilon_{2n}}{R} \right)^3}{\rho_1\coth(\frac{\epsilon_{2n}}{R}h_1) + \rho_2\coth(\frac{\epsilon_{2n}}{R}h_2)} \label{eq:Dispersion}
\end{equation}
are the natural eigenfrequencies of gravity-capillary waves in cylinders and
\begin{equation}
 F_{2n} = \frac{2(\sigma_1 - \sigma_2) B_0^2 \Omega^2 R^2}{\left[\rho_1\coth(\frac{\epsilon_{2n}}{R}h_1) + \rho_2\coth(\frac{\epsilon_{2n}}{R}h_2)\right](\epsilon_{2n}^2 - 4)J_2 (\epsilon_{2n})}
\end{equation}
can be considered as mode-dependent forcing parameters.  
This way, we have reduced a set of four partial differential equations (\ref{Eq:a} - \ref{Eq:b}) together with five boundary conditions (\ref{Eq:c} - \ref{Eq:g}) into an infinite set of decoupled ordinary differential equations. Stationary solutions of the modal equations (\ref{Eq:ModalA}) and (\ref{Eq:ModalB}) can be obtained as follows.
\begin{align}
	\alpha_{2n}(t) = \frac{F_{2n}}{4\Omega^2 - \omega_{2n}^2}\cos(2\Omega t), \ \ \beta_{2n}(t) = \frac{F_{2n}}{4\Omega^2 - \omega_{2n}^2}\sin(2\Omega t). \label{Eq:ModalSol}
\end{align} 
Substituting the solutions (\ref{Eq:ModalSol}) back into the ansatz potentials (\ref{eq:Pot1}) and (\ref{eq:Pot2}) finally yields the forced potentials solutions
\begin{align}
	\phi_{1}(r,\theta,z,t) &= - \sum_{n=1}^{\infty}\frac{2(\sigma_1 - \sigma_2) B_0^2 R^3 \Omega^2 \omega_{2n}^2}{\left[(\rho_2 - \rho_1)g + \gamma \frac{\epsilon_{2n}^2}{R^2}\right](4\Omega^2 -\omega_{2n}^2)}\frac{\cosh \left(\frac{\epsilon_{2n}}{R}(z-h_1)\right)}{\sinh \left(\frac{\epsilon_{2n}}{R} h_1 \right)} \nonumber \\
	&\times \frac{J_2 \left(\frac{\epsilon_{2n}r}{R}\right)}{\epsilon_{2n}(\epsilon_{2n}^2 - 4) J_2 \left(\epsilon_{2n}\right)}\cos(2\Omega t - 2\theta), \label{eq:PotSol1} \\
	\phi_{2}(r,\theta,z,t) &= \sum_{n=1}^{\infty}\frac{2(\sigma_1 - \sigma_2) B_0^2 R^3 \Omega^2 \omega_{2n}^2}{\left[(\rho_2 - \rho_1)g + \gamma \frac{\epsilon_{2n}^2}{R^2}\right](4\Omega^2 -\omega_{2n}^2)}\frac{\cosh \left(\frac{\epsilon_{2n}}{R}(z+ h_2)\right)}{\sinh \left(\frac{\epsilon_{2n}}{R} h_2 \right)} \nonumber \\
	&\times \frac{J_2 \left(\frac{\epsilon_{2n}r}{R}\right)}{\epsilon_{2n}(\epsilon_{2n}^2 - 4) J_2 \left(\epsilon_{2n}\right)}\cos(2\Omega t - 2\theta).  \label{eq:PotSol2} 
\end{align}
The corresponding interface elevation $\eta (r,\theta,t)$ derives from the boundary condition (\ref{Eq:f}) and can be stated as
\begin{align}
\eta(r,\theta ,t) = \sum_{n=1}^{\infty} \frac{(\sigma_2 - \sigma_1) B_0^2 R^2 \Omega \omega_{2n}^2}{\left[(\rho_2 - \rho_1)g + \gamma \frac{\epsilon_{2n}^2}{R^2}\right](\omega_{2n}^2 - 4\Omega^2)  }\frac{J_2\left(\frac{\epsilon_{2n}r}{R}\right)}{(\epsilon_{2n}^2 - 4)J_2(\epsilon_{2n})}\sin(2\Omega t - 2\theta). \label{Eq:Elev}
\end{align}
The wave elevation grows with the square of the magnetic field $\sim B_0^2$ and tank radius $\sim R^2$, which is why waves tend to be critical particularly in large-scale stirrers operating in the range of $B_0 \sim 0.1\, {\rm T}$. Interestingly, the solution predicts that waves cannot occur for $\sigma_1 = \sigma_2$. The reason behind this behavior is that the induced magnetic pressure is equal in both layers so that no net force is acting on the interface within the framework of our idealized description. This is not necessarily true anymore if we had included horizontal wall effects, but even then resulting wave motions would be vanishingly small in cases where the fluid layers have comparable heights. We can conclude that significant wave motions are only to be expected if one pairs a highly conducting fluid with a poorly conducting fluid, as it is the case in the most relevant application of free liquid metal surfaces. It can further be seen that the sign change appearing in solution (\ref{Eq:Elev}) between the two cases $\sigma_1 > \sigma_2$ and $\sigma_1 < \sigma_2$ causes a phase shift of $90^{\circ}$ because $\sin(2\Omega t - 2\cdot90^{\circ}) = -\sin(2\Omega t - 2\cdot0^{\circ})$. The simple explanation is (as long as we remain below the first resonance frequency $\Omega < \omega_{21}/2$, see next section) that a force field locally pointing towards the side wall below the interface leads to a local heightening and the same force located above the interface leads to a local lowering of the interface.
\subsection{Theoretical results}
In the following, we will elucidate the characteristics of the wave solution and discuss its underlying physics in more detail. For the sake of simplicity, we rewrite Eq.\,(\ref{Eq:Elev}) in a dimensionless from by introducing the dimensionless variables $\tilde{r} = r/R, \ \tilde{z} = z/R$ and $\tilde{t} = \Omega t$, yielding
\begin{align}
	\frac{\eta (\tilde{r},\theta,\tilde{t})}{R}=  \sum_{n=1}^{\infty}\frac{{\rm sgn}(\sigma_2 - \sigma_1)Fr}{\left(1 + \frac{\epsilon_{2n}^2}{Bo}\right)}\frac{\Gamma_{2n}^2}{\Gamma_{2n}^2 - 4}\frac{J_2\left(\epsilon_{2n}\tilde{r}\right)}{(\epsilon_{2n}^2 - 4)J_2(\epsilon_{2n})}\sin(2\tilde{t} - 2\theta),\label{Eq:ElevNonDim} \\
	{\rm with} \  \Gamma_{2n}^2 = \frac{\omega_{2n}^2}{\Omega^2} = \frac{N_2 - N_1}{Fr}\frac{2A\epsilon_{2n}\left(1+\frac{\epsilon_{2n}^2}{B_0}\right)}{(1-A)\coth(\epsilon_{2n}H_1) + (1+A)\coth(\epsilon_{2n}H_2)}.
\end{align}
The magnetic Froude number $Fr$ is the key driving parameter governing non-resonant wave motions, the Bond number $Bo$ is likewise significant as it indicates the transition into the capillary wave regime. This is important because it shows that the excited waves do not contain infinitely many length scales, as it appears to be the case through the infinite sum in solution (\ref{Eq:ElevNonDim}). The solution actually converges very fast since the term $1+\epsilon_{2n}^2 /Bo$ appearing in the denominator of solution (\ref{Eq:ElevNonDim}) increases for any given $Bo$ with the wave numbers $\epsilon_{2n}$, so that summands become increasingly smaller and higher wave modes are finally damped out by interfacial tension at the point where wave lengths fall considerably below the capillary length $l_{\rm cap} = R/\sqrt{Bo}$. Further, we see that all modes diverge at the frequencies $\Gamma_{2n}^2 = 4 \Leftrightarrow \Omega = \pm  \omega_{2n}/2$, determining the resonance conditions. Resonance occurs at half the  eigenfrequency of the wave modes $(2,n)$ and responding waves always rotate with twice the frequency of the applied rotating magnetic field, i.e., waves always follow the induced rotating Lorentz force potential (\ref{eq:PotL}). This statement, however, is only true for linear wave responses. For example, super-harmonic waves can be excited at fractions of the natural frequencies \citep{Reclari2014,Bongarzone2022} under weakly nonlinear forcing conditions. The singularities at resonances in solution (\ref{Eq:ElevNonDim}) occur here only as an artifact from having neglected dissipation in our wave model. The wave solution is therefore non-physical in the close vicinity of the eigenfrequencies $\omega_{2n}$, amplitudes must stay finite in reality. This problem is usually circumvented by equipping the modal equations (\ref{Eq:ModalB}) with (linear) damping parameters, which can close the resonance curves, see \cite{Horstmann2020}. Viscous damping rates that can be calculated from Stokes boundary layers developing at the tank walls (and above and below the interfaces in the case of two-fluid stratifications) are well known for free-surface \cite{Case1957} and interfacial \cite{Herreman2019a} waves in upright circular cylinders. In our case, however, magnetic damping accounts for a significant part of the total dissipation. 
Magnetic damping appears here because the waves rotate with twice the frequency of the magnetic field $2\Omega$, which is equivalent to a wave rotating with $\Omega$ through a static magnetic field. Whenever an electrically conducting fluid moves through a magnetic field, Lenz's law requires that a Lorentz force is induced which exactly opposes its causative motion. This magnetic damping contribution is, however, rather difficult to calculate and due to the superimposed swirling flow driven by the mean part of the Lorentz force (\ref{eq:Lorentz}), there is a third source of dissipation\,``Ekman pumping''\cite{Davidson1992,Davidson1995} present in the system. Due to these intricacies, we restrict the model to dissipationless non-resonant wave excitations.
\begin{figure}
	\hspace*{-0.5cm}
	\centering
	\includegraphics[width=1.04\columnwidth]{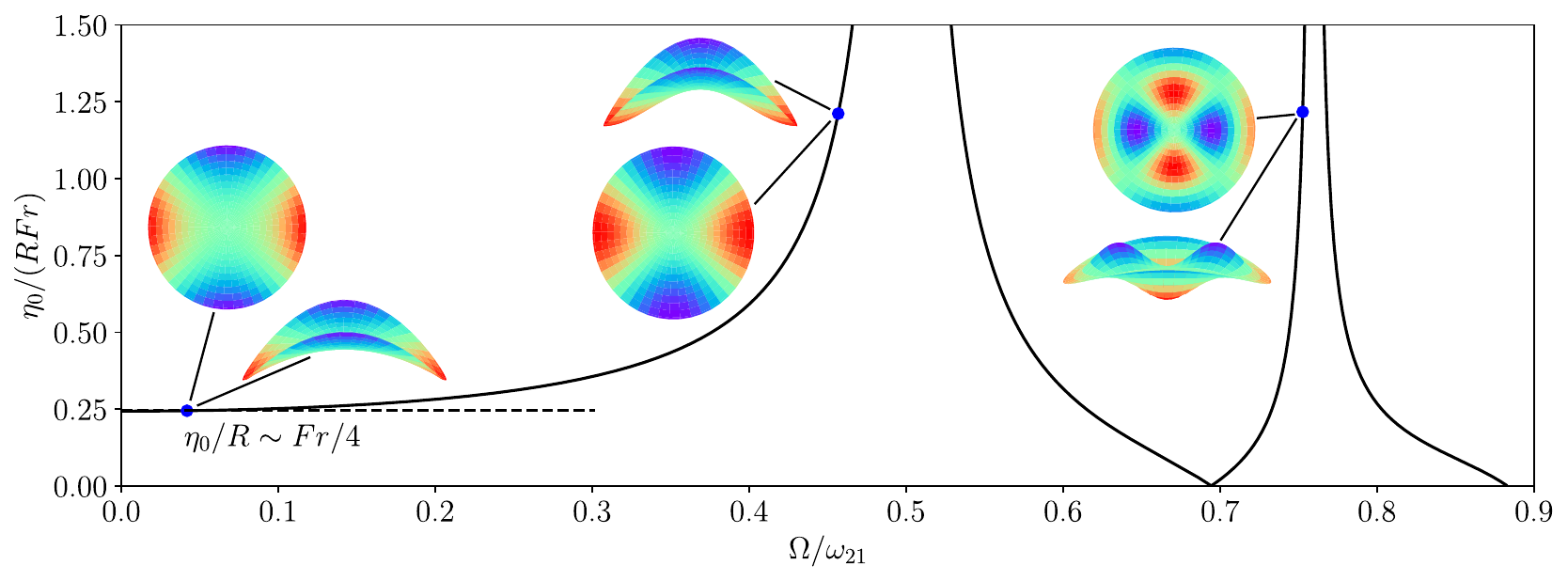}
	\caption{Maximum wave elevation at the tank wall $\eta_0 = \eta(r=R)$ normalized by the container radius $R$ and the Froude number $Fr$ as a function of the excitation frequency $\Omega$ normalized by the first natural eigenfrequency $\omega_{21}$. Additionally, normalized 3D visualizations of excited interface elevations are shown at three different points in the linear regime and close to the first and second resonance from two different perspectives.  \label{f:ModeCurve}}
\end{figure}

In order to gain further insight into the excited wave dynamics, we show in Fig.\,\ref{f:ModeCurve} maximal wave elevations (normalized by the tank radius $R$ and the Froude number $Fr$) as a function of the RMF excitation frequency $\Omega$ (normalized by the first natural eigenfrequency $\omega_{21}$) in the limit of gravity waves $Bo \longrightarrow \infty$. Different wave regimes become evident, which are characterized by different wave forms also visualized in Fig.\,\ref{f:ModeCurve}. For small driving frequencies far before the first resonance $\Omega \ll \omega_{21}/2$, wave amplitudes grow linearly with the magnetic Froude number $\eta_0 \sim Fr$ and the interface takes the exact shape of a hyperbolic paraboloid. This is formally not a wave solution, the interface displacement follows here directly from the balance between the Lorentz force and the restoring gravitational acceleration. This can be shown by applying the limit $Bo \longrightarrow \infty$ to solution (\ref{Eq:ElevNonDim}) and rearranging the resonance terms as 
\begin{equation}
\frac{\Gamma_{2n}^2}{\Gamma_{2n}^2 - 4} = 1 + \frac{4}{\Gamma_{2n}^2 - 4},
\end{equation}
giving
\begin{align}
	\frac{\eta (\tilde{r},\theta,\tilde{t})}{R}=  {\rm sgn}(\sigma_2 - \sigma_1)Fr\sum_{n=1}^{\infty}\left(1 + \frac{4}{\Gamma_{2n}^2 - 4}\right)\frac{J_2\left(\epsilon_{2n}\tilde{r}\right)}{(\epsilon_{2n}^2 - 4)J_2(\epsilon_{2n})}\sin(2\tilde{t} - 2\theta).
\end{align}
Now, the Bessel series (\ref{eq:Bessel}) can be reinserted to eliminate the frequency-independent part of the series, finally yielding
\begin{align}
	\frac{\eta (\tilde{r},\theta,\tilde{t})}{R}= \frac{{\rm sgn}(\sigma_2 - \sigma_1)Fr}{4} \left[\tilde{r}^2 + \sum_{n=1}^{\infty}\frac{16}{\Gamma_{2n}^2 - 4}\frac{J_2\left(\epsilon_{2n}\tilde{r}\right)}{(\epsilon_{2n}^2 - 4)J_2(\epsilon_{2n})}\right]\sin(2\tilde{t} - 2\theta)
\end{align}
For sufficiently small RMF frequencies $\Omega \ll \omega_{21}/2$, the first parabolic term dominates the solution. In Cartesian coordinates it reads
\begin{align}
	\frac{\eta (x,y,t)}{R} \approx \frac{{\rm sgn}(\sigma_2 - \sigma_1)Fr}{4R^2}\left[(x^2 - y^2)\cos(2\Omega t) + 2xy\sin(2\Omega t)\right] \ {\rm for} \ x^2 + y^2 < R^2,
\end{align}
which is the normal form of a hyperbolic paraboloid rotating with frequency $2\Omega$. This solution is the counterpart to the rotating or oscillating disc (or plane) solution appearing in many sloshing problems \cite{Horstmann2020} under low-frequency excitation. However, it should be mentioned that this solution violates the beforehand assumed static contact angle condition, so that this interface pattern can only emerge in sufficiently large vessels $R \gg l_{\rm cap}$. In the case of small Bond numbers $Bo \lesssim 1$, all summands of the series (\ref{Eq:ElevNonDim}) are wave number dependent and the hyperbolic paraboloid solution does not exist anymore. The interface has then the shape of Bessel functions. The same transition takes place, independently of $Bo$, when $\Omega$ approaches the first resonance condition $\Omega = \omega_{21}/2$. The radial shaping is here predominantly described by the Bessel function of second order $\sim J_2 (\epsilon_{21}\tilde{r})$, which now everywhere maintains the static contact angle condition of $90^{\circ}$. This is a true wave solution coming along with all common wave properties, as, in particular and in contrast to the hyperbolic paraboloid solution, it drives secondary flows resulting in an azimuthal mean mass transport commonly referred to as the ``Stokes drift'' \cite{Bouvard2017}. Once the RMF frequency exceed the resonance condition $\Omega > \omega_{21}/2$, the solution (\ref{Eq:ElevNonDim}) changes its sign and the wave undergoes a phase shift of $90^{\circ}$. The interface elevation at the side wall is now locally in opposition to the Lorentz force distribution, similar to the cases of better-known $180^{\circ}$ phase jumps appearing in $m=1$ sloshing problems around resonance. The second resonance $\Omega = \omega_{22}/2$ is also shown in Fig.\,\ref{f:ModeCurve}, which is distinguished by the appearance of $n=2$ antinodal cycles. Compared to the first resonance, the second resonance arises in a significantly reduced frequency range, which is also true for all higher resonances. Therefore, the first wave mode can be regarded as the dominant mode in magnetic stirring, reaching the highest amplitudes in practice. Whenever a two-layer system is stirred with an RMF frequency close to $\omega_{21}/2$, the possible occurrence of critical free-surface or interface displacements should be taken into account.      

\section{Experiments}
\label{sec:Exp}
\begin{figure}
	\centering
	\def\svgwidth{465pt}    
	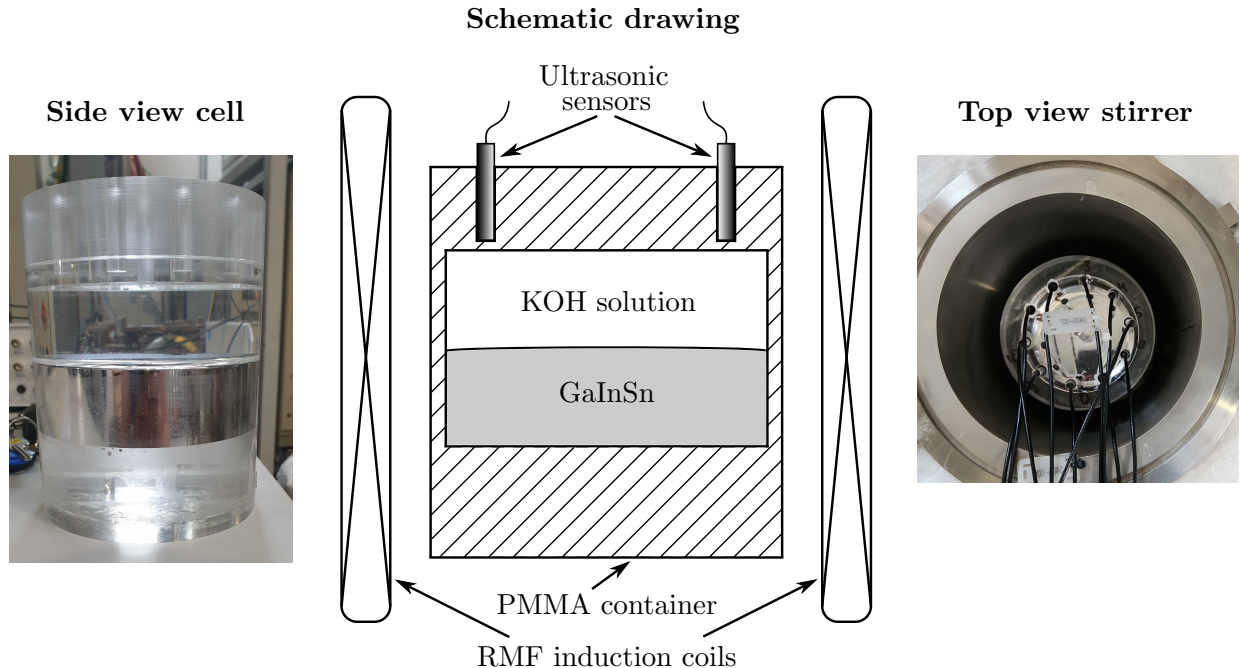	
	\caption{Sketch and photographs of the experimental setup.}
	\label{fig:ExpSetup}
\end{figure}
\subsection{Experimental setup}
The experiments were conducted in an upright cylindrical container ($R=5\, {\rm cm}, h=7.5\, {\rm cm}$) made of polished poly(methyl methacrylate)
(PMMA) that is filled with two immiscible liquids self-stratifying due to gravity, see Fig.\,\ref{fig:ExpSetup}. The room temperature-liquid eutectic alloy GaInSn is filling the bottom space and functions here as the MHD active liquid in which Lorentz forces are induced under exposure to magnetic fields. It has a kinematic viscosity of $\nu = 3.4\times 10^{-7}\, {\rm m}^2\,{\rm s}^{-1}$, a density of $\rho = 6.36\times10^3\,{\rm kg}\, {\rm m}^{-3}$ and a high electrical conductivity of $\sigma = 3.2\times 10^6\,{\rm S}\,{\rm m}^{-1}$ at lab temperature $T=20^{\circ}{\rm C}$. An aqueous $1\, \%$ caustic potash (KOH) solution is layered on top and has virtually the same material properties of water. Caustic potash was chosen for two reasons. First and foremost, we intended to avoid the formation of an oxide layer at the metal surface, which immediately develops once GaInSn comes into contact with air; and it also cannot be avoided when GaInSn is in contact with most liquids such as water and oils. The oxide layer is known to have elastic properties (similar to a membrane floating on top of a liquid \cite{Zhang2011}), causes additional dissipation and complicates the contact line boundary conditions. For this reason, we wanted to prevent the alloy from oxidizing in line with the objectives of this study, however, the extension of our model by elastic wave motions might be rewarding for a better understanding of RMF-induced oxide layer motions being a determinant in various practical stirring applications. The oxide layer can easily be neutralized by layering different alkaline and acidic solutions, e.g., hydrochloric acid (HCl) was employed in many studies \cite{Travnikov2012,Vogt2013,Vogt2015,Burguete2012,Grants2021a}. 
However, in preliminary experiments we observed that HCl caused a pronounced meniscus not satisfying the contact line boundary conditions of our wave model. We found that KOH remedies this issue, it forms only a very small meniscus of length $\sim 1\, {\rm mm} \ll R = 5\, {\rm cm}$, see Fig.\,\ref{fig:ExpSetup}, and allows the contact line to slide freely along the side wall (no pinning effects). 

The internal dimensions of the cylindrical cell measure a radius of $R=5\, {\rm cm}$ and a total height of $h_1 + h_2 = 7.5\, {\rm cm}$, in which the interface position was varied during the experiments. The upper lid covers ten equally distributed sockets for the attachment of ultrasonic sensors. The distance between the rotational axis of the cylinder and the center of the sockets is $4.2\, {\rm cm}$, sufficiently far away from the side wall to prevent acoustic interferences caused by wall reflections. To ensure non-invasive measurements, the ultrasonic probes are
not in direct contact with the working liquids, a $5\, {\rm mm}$ thick
base was kept between the probes and the interior space.
The entire cell is placed concentrically in the
$20\, {\rm cm}$ bore hole of the magnetic induction system PERM available
at HZDR, which has already been used in a number of studies \cite{Grants2008,Denisov2010,Vogt2012,Vogt2013}. The system can both induce traveling and rotating magnetic fields. The RMF used in this study is generated by a three-phase current applied to 
a radial arrangement of six induction coils, in which opposing coils are connected as pole pairs, analogously to the typical construction of a stator in asynchronous motors. This configuration induces a magnetic field vector rotating homogeneously (over a height span of $\sim 20\, {\rm cm}$ fully encompassing the cell) in the horizontal plane as described by Eq.\,(\ref{eq:RMF}).  We applied small RMF frequencies $f = \Omega/2\pi = 1$-$10\, {Hz}$ (to meet the first wave's natural frequencies) and magnetic fields up to $B_0 = 5\, {\rm mT}$. These values correspond to a minimal skin depth of $\delta_s = 8.8\, {\rm cm}$, which is larger than the cell radius and thereby guarantees that the liquid metal is always fully immersed in the RMF. Finally, it must be noted that the PERM setup generates a rotating magnetic field in clockwise direction. The wave model was formulated for an RMF with a mathematically positive (counterclockwise) sense of rotation, but due to the time-reversal invariance fulfilled for all governing equation, the solutions can still be applied one-to-one to the experiments only by assigning $\Omega \rightarrow -\Omega$. 
\subsection{Ultrasound measurements}
In order to measure the interfacial motions, we use the ultrasound pulse-echo technique presented in \citet{Horstmann2019} that was specially devised to detect wave motions in opaque liquids. This method is based on Ultrasound Doppler Velocimetry (UDV), a widespread technique established for high-resolution velocity measurements in liquid metals. The original measuring method relies on the Doppler shift. UDV probes emit consecutively short ultrasonic pulses, which are reflected on existing or artificially added scattering particles in the working liquid. After some transit time depending
on the distance of the single scattering particles, the pulse echoes are recaptured by the probes. By means of differences in transit time of the scattering echoes between consecutive pulse emissions, resulting from a finite displacement of the particle position locally following the velocity field, the axial velocity component (in direction of the pulse) can be obtained. However, as demonstrated in \cite{Horstmann2019}, this technique can likewise be utilized for the further purpose of measuring position and movements of free surfaces and interfaces. The idea is to track the echo signal corresponding to the interface directly. Here one encounters the problem that the echo is not maintaining its shape throughout interface movements, because changes in the orientation of the interface cause part of the echo signal to be reflected away from the UDV probe. This limits the largest measurable interfacial elevations ($\eta_{\rm max}/R \lesssim 20\, \%$), however, vertical interface motions can still be measured reliably when one tracks the first echo value of the beginning echo signal curve that is significantly higher than the echo values of the noise level, see \cite{Horstmann2019}. Exactly this method was reused for this study. We employed the ultrasound Doppler velocimeter DOP 3010 from Signal Processing, which can drive up to 10 ultrasound probes we adjusted into the top lid of the cell, see Fig.\,\ref{fig:ExpSetup}. The utilized ultrasound probes operate with $4\, {\rm MHz}$ and encompass piezoelectric transducers of $5\, {\rm mm}$ diameter. All probes were lubricated with ultrasonic gel before we inserted them into the cylinder to optimize the acoustic coupling with the PMMA. Due to the significantly better acoustic reflection capacity of the GaInSn-KOH interface compared to water-oil interfaces, we were able to achieve even better resolution than in \cite{Horstmann2019}, allowing us to measure wave amplitudes as low as up to $\eta_0 \sim 0.01\, {\rm mm}$.
\begin{figure}
	\vspace*{-0.5cm}
	\hspace*{-0.5cm}
	\centering
	\includegraphics[width=1.03\columnwidth]{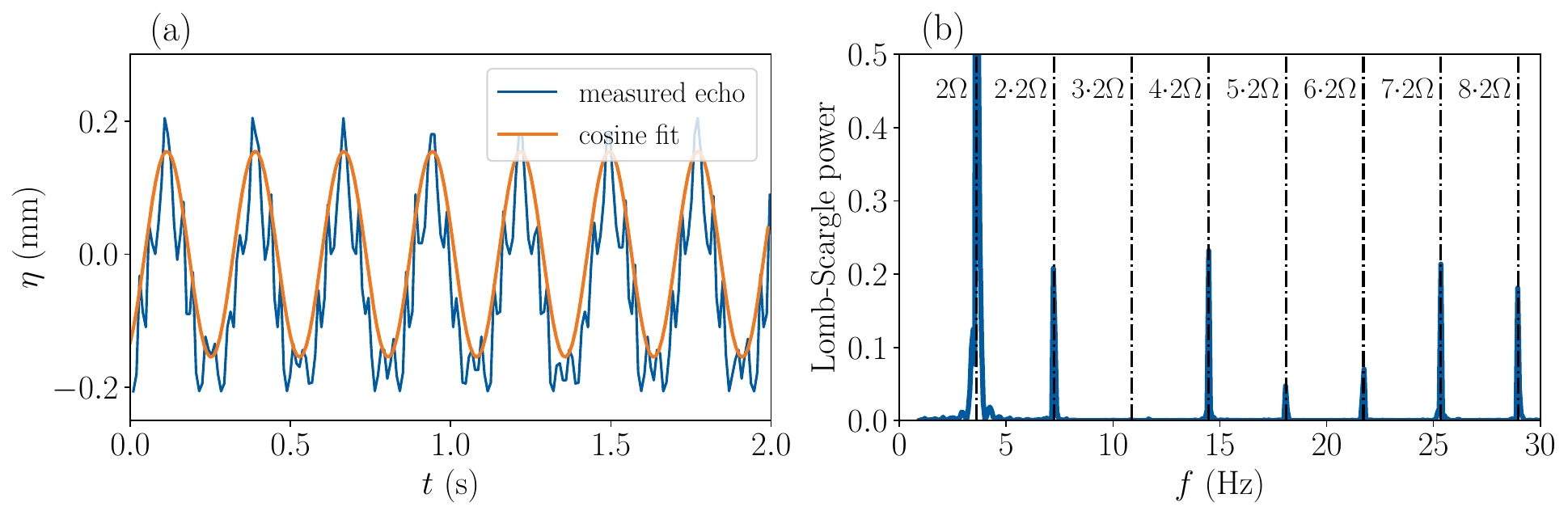}
	\caption{(a) Measured local interfacial echo distance $\eta$ of sensor 1 and fitted cosine function as a function of time $t$ for the applied RMF frequency $f=\Omega/(2\pi) = 1.81\, {\rm Hz}$ and magnetic field $B_0 = 4.5\, {\rm mT}$. (b) Lomb-Scargle spectrum corresponding to the measured echo signal.  \label{f:Signal}}
\end{figure}

To demonstrate our approach to signal processing, 
Fig.\,\ref{f:Signal}\,(a) shows an example of a recorded single echo signal corresponding to resonantly excited wave motion $\Omega = \omega_{21}/2$. A harmonic time response is clearly recognizable, which is, however, overshadowed by several higher frequencies. This was initially surprising, as we would have expected an undisturbed harmonic wave at such small amplitudes, as it usually occurs in linear sloshing. But higher frequencies were always contained in the  measured echo signals, also in all non-resonant cases. For this reason, we have analyzed the underlying frequency spectra of all conducted measurements and always found frequency patterns similar to Fig.\,\ref{f:Signal}\,(b). As expected, the clearest peak is evident at the resonance frequency $2\Omega$. Interestingly, several more peaks can be noticed at higher frequencies, which are always multiples of twice the RMF frequency $2\Omega$. These are obviously higher harmonics of the excited fundamental frequency. In Sec.\,\ref{Sec:Harmonics} we will reveal why higher harmonics are inevitably excited by the Lorentz force even in the small-amplitude limit. As part of signal processing, the presence of higher harmonics impact the accuracy of the amplitude evaluation, because maximum and minimum interface displacements can no longer be unambiguously assigned to the primary irrotational wave motion described by Eq.\,(\ref{Eq:ElevNonDim}). 
\begin{figure}
	\centering
	\vspace*{-0.6cm}
	\hspace*{-0.6cm}
	\def\svgwidth{490pt}    
	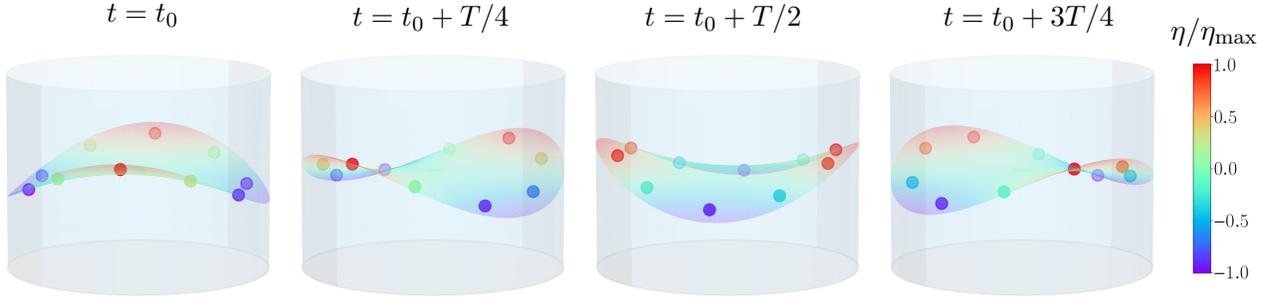
	\vspace*{-2.cm}	
	\caption{3D visualization of a magnetically excited interface at four different times within one period $T$ on the basis of ten simultaneously applied ultrasound probes. The measured interface elevation $\eta/\eta_{\rm max}$ of all probes are coded in color. Further, surface fits of the wave mode $(2,1)$ are included and also color-coded.}
	\label{f:3D}
\end{figure}

To be further able to reliably identify the amplitude corresponding to the primary wave response, we always fitted sinusoidal signals to the echo data, see Fig.\,\ref{f:Signal}\,(a). We always took the fitted sinusoidal waveform as the basis for determining the local primary amplitudes at the UDV sensor positions $r=0.42\, {\rm cm}$ to be discussed in the following chapters. Through the simultaneous operation of up to ten UDV probes, we can further resolve excited wave motion spatially. Fig.\,\ref{f:3D} shows the reconstructed three-dimensional interface elevation $\eta$ corresponding to the sample measurement of Fig.\,\ref{f:Signal}. To this end, we have visualized the fitted interface positions of all ten UDV probes for four chosen times within one wave period $T = \pi/\Omega$. Both the $m=2$ paraboloidal waveform predicted by the solution (\ref{Eq:ElevNonDim}) and a clockwise rotational motion become clearly visible.
The comparison with the surface fits of solution (\ref{Eq:ElevNonDim}), which are also displayed in Fig.\,\ref{f:3D}, confirms the existence of the predicted irrotational wave created upon magnetic stirring. It is important to note, however, that the radial shaping of wave modes cannot be resolved due to the circumferential arrangement of the ten ultrasonic probes. Therefore, the non-resonant hyperbolic paraboloid and the resonant Bessel wave states predicted in the first two waves regimes (see Fig.\,\ref{f:ModeCurve}) are indistinguishable in our experiment and cannot be independently verified on the basis of such qualitative benchmarks. In order to verify the significance and applicability of our simplified potential flow model to the quite complex and multifaceted RMF-driven wave flow, amplitude curves and resonance frequencies are quantitatively compared against the theoretical predictions in the following sections.  

\section{Results}
\label{Sec:Results}
\subsection{General resonance profiles}
\begin{figure}
	\hspace*{-0.5cm}
	\centering
	\includegraphics[width=1.04\columnwidth]{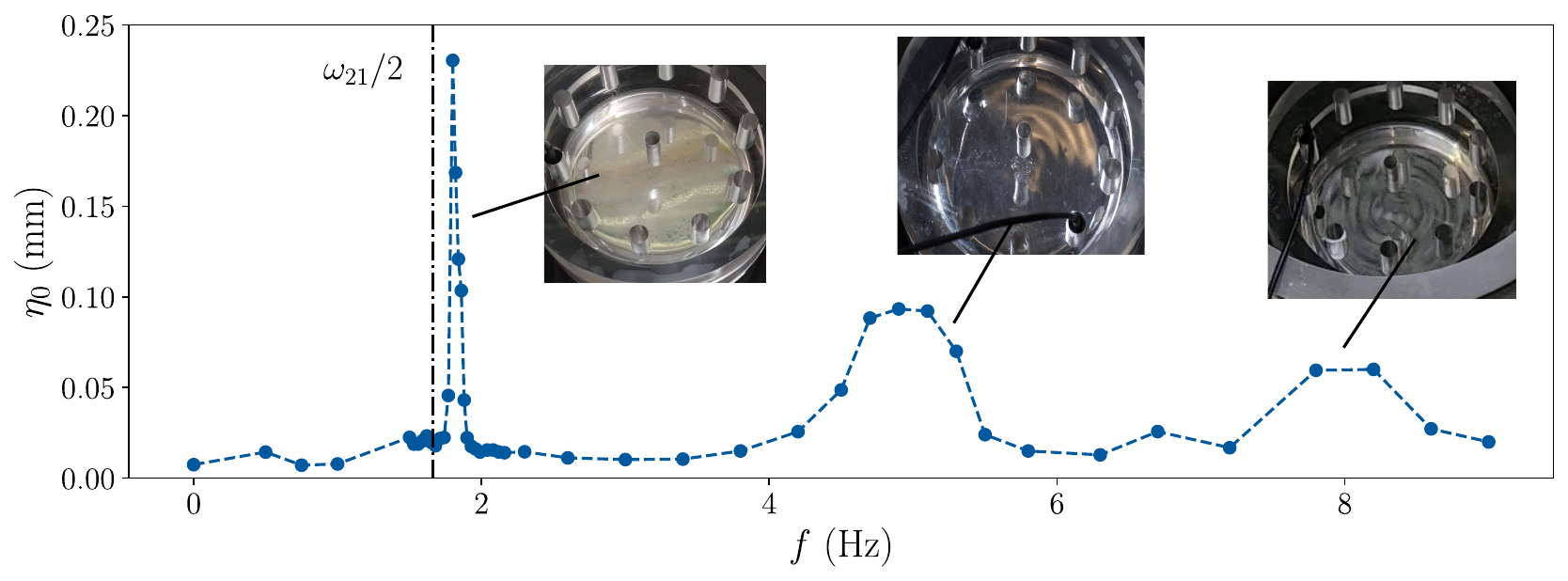}
	\caption{Measured wave amplitude $\eta_0$ at the sensor position $\eta(r=4.2\,{\rm mm})$ as a function of the RMF frequency $f=\Omega/(2\pi)$ for an applied magnetic field of $B_0 = 4.6\, {\rm mT}$ and layer heights of $h_1 = 4\, {\rm cm}$ and $h_2 = 3.5\, {\rm cm}$. Moreover, snapshots of the metal surface are included to emphasize different observed wave regimes. \label{f:ExampleCurve}}
\end{figure}

For our chosen stratification of GaInSn and KOH with layer thicknesses $h_1 = 4$\textendash$4.5\,{\rm cm}$ and $h_2 = 3.5$\textendash$3\,{\rm cm}$ and estimated interfacial tension of $\gamma \approx 0.6\,{\rm N}\,{\rm m}^{-1}$,  Eq.\,(\ref{eq:Dispersion}) predicts the first and dominant resonance to occur around an applied RMF frequency of $f = \omega_{21}/(4\pi) = 1.66$\textendash$1.68\,{\rm Hz}$. Consequently, we limited ourselves to slow RMF frequencies and carried out measurements mostly in the range from 0 to $10\,{\rm Hz}$. A typical profile of the observed amplitude-frequency curves within this range is depicted in Fig.\,\ref{f:ExampleCurve}. The profile comprises different regimes, to which we have included representative snapshots of the vibrating liquid metal interface. The wave motions cannot be properly recognized on the photos, we therefore refer the interested readers to the accompanying videos available in the supplementary material. 

At all excitation frequencies, small displacements of the metal surfaces were visible in form of reflection patterns. For frequencies up to around $\lesssim 3\,{\rm Hz}$, the surface remained largely smooth and followed the movement pattern of the anticipated rotating hyperbolic paraboloid. As predicted, we always observed sharp peaks in the amplitude near the resonance frequency $f = \omega_{21}/(4\pi)$, however, the measured resonance frequencies were always found to be up to $10\,\%$ higher than the predicted ones. The more we increased the magnetic field $B_0$, the higher was the deviation between the predicted and measured frequencies. This behavior gives rise to a noticeable frequency shift between the theoretical and experimental resonance curves. We will address this issue in the next section. For higher RMF frequencies $f \gtrsim 3\,{\rm Hz}$, amplitude responses were not reproducible and we have observed the appearance of different higher wave mode patterns resembling Faraday waves. There are still regions with higher amplitude responses at around $5\,{\rm Hz}$ and $8\,{\rm Hz}$ in this sample measurement. However, these do not represent classic resonances because the amplitude profiles have been observed to change significantly with $B_0$. Resonances at higher modes $\omega_{22}/(4\pi), \omega_{23}/(4\pi), \omega_{24}/(4\pi), ...$ were also not visible, which is, however, explained by the arrangement of the UDV sensors. Wave elevations are only measured locally at radial positions close to the side wall ($r=4.2\, {\rm cm}$). But higher modes form anti-nodal circles farther inwards, see Fig.\,\ref{f:ModeCurve}, and are therefore mostly outside the spatial measuring window.  For this reason, we can only capture the first $\omega_{21}$ resonance quantitatively and only properly compare measured amplitudes with the linear wave model around the first observed peak, which is why we will focus on this regime in the following.   

\subsection{Frequency correction}
Before wave responses can be adequately compared to the wave model, we have to resolve the observed frequency shift between measurements and the natural resonance frequency. The discrepancy stems from the fact that we only considered the irrotational part of the Lorentz force driving the wave motion. But there is also the mean part of the Lorentz force, Eq.\,(\ref{Eq:FMean}), driving a constant azimuthal swirling flow. This means that the RMF always excites a superimposed state of oscillatory wave motion and a mean swirling flow, where the latter causes the frequency shift. It can be understood as follows: the wave always follows the rotational motion of the induced Lorentz force synchronously and thereby rotates with angular frequency $2\Omega$ from the view of the lab's frame of references. However, the carrier media (GaInSn and KOH) of the interfacial waves itself are rotating in the same direction as the wave but with lower angular frequency $v(r) < r 2\Omega$. In relation to the co-rotating carrier media the waves propagate respectively slower. Consequently, the resonant frequencies measured in the stationary laboratory must be higher than the theoretical resonance frequency $f = \omega_{21}/(4\pi)$, which only apply in the inertial frame of reference. Keeping this in mind, we can easily estimate the observed frequency shift simply by calculating the azimuthal swirling velocity induced by the mean part (Eq.\,(\ref{Eq:FMean})) of the Lorentz force. For the sake of simplicity, we again neglect the influence of the upper and lower ends of the cylinder. Then, the azimuthal velocity profile directly derives from the balance between the electromagnetic and viscous forces and can be stated in terms of the Hartmann number $Ha$ as follows \cite{Tagawa2019}:     
\begin{align}
	v(r) = r\Omega - R\Omega\frac{I_1 \left(\frac{Ha\cdot r}{\sqrt{2}R}\right)}{I_1 \left(Ha / \sqrt{2}\right)} \ \ {\rm with} \ \ Ha = B_0 R \sqrt{\frac{\sigma_2}{\rho_2 \nu_2}}.\label{Eq:VelProfile}
\end{align}  
Here, $I_1 (x)$ denotes the modified Bessel function of the first kind and first order. The solution comprises two parts. The first part $r\Omega$ describes a solid body rotation, which is occupied by the liquid in the center of the tank (small radii). In this domain the liquid follows the mean Lorentz force and no energy is injected. The swirling motion is instead driven in the peripheral region near the side wall, where the azimuthal velocity drops to zero in order to fulfill the no-slip boundary condition. This drop is described by the Bessel function term of Eq.\,(\ref{Eq:VelProfile}) and is restricted to close proximity of the side wall in the case of higher Hartmann numbers $Ha \gtrsim 20$.

The solution, however, is formally only valid for single-liquid systems. In our experiment, the flow is excited solely in the lower liquid metal, but the upper KOH layer is entrained and also follows the rotational motion. Additional friction losses are created in the upper layer, in which no energy is injected. Due to this difference and the presence of vertical walls, there are significantly higher frictional losses than those captured by solution (\ref{Eq:VelProfile}). We therefore correct the flow by an empirical factor $\chi$. The frequency associated to the swirl flow evaluated at the UDV sensor position $r=r_0 =4.2\,{\rm cm}$ is then calculated as 
\begin{align}
	f_{\rm swirl}(r_0) = \chi\frac{1}{2\pi}\left[\Omega - \frac{R\Omega}{r_0}\frac{I_1 \left(\frac{Ha\cdot r_0}{\sqrt{2}R}\right)}{I_1 \left(\frac{Ha}{\sqrt{2}}\right)}\right]. \label{Eq:SwirlFreq}
\end{align}    
\begin{figure}
	\hspace*{-0.5cm}
	\centering
	\includegraphics[width=0.8\columnwidth]{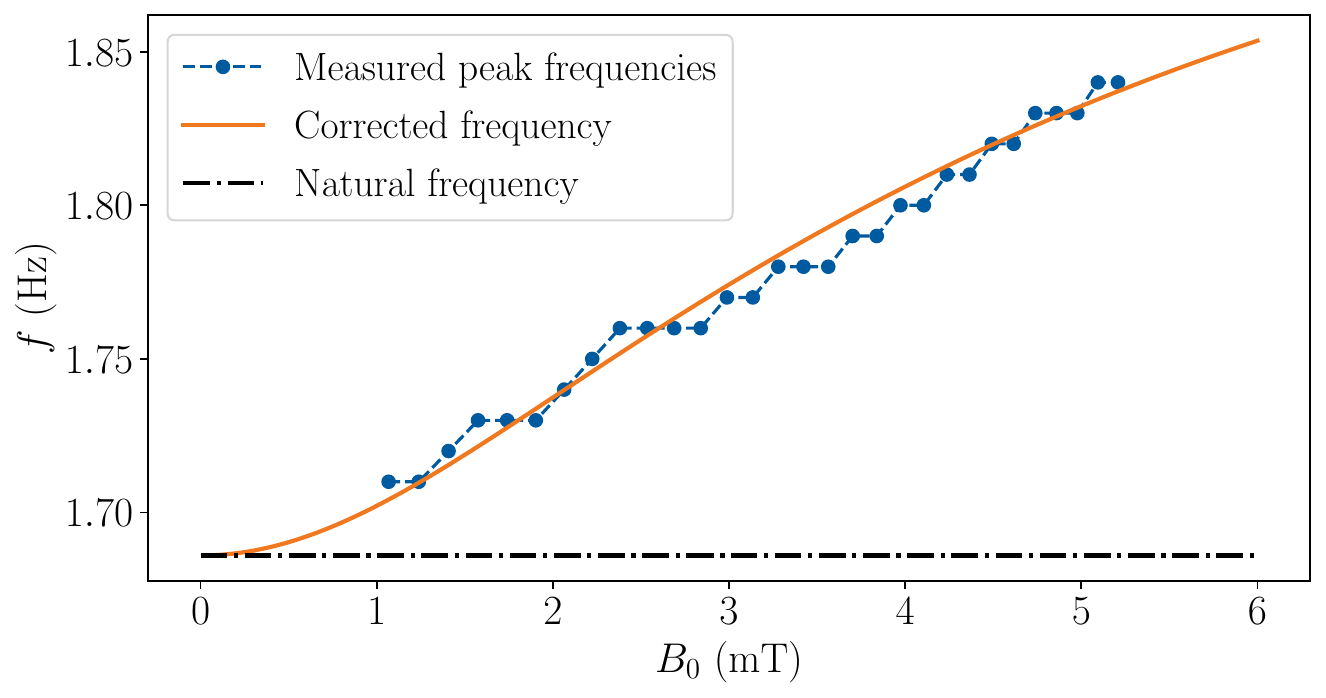}
	\caption{Measured peak frequencies $f$ and theoretical estimates as a function of the applied magnetic field $B_0$. The orange line shows corrected frequencies including swirling frequencies with the correction factor $\chi=0.15$ and the natural resonance frequency $f = \omega_{21}/(4\pi)$ is shown by the dotted black line. \label{f:MagneticFrequency}}
\end{figure}

Fig.\,\ref{f:MagneticFrequency} shows the sum of the magnetic swirling frequency $f_{\rm swirl}$ with $\chi=0.15$ and natural resonance frequency $\omega_{21}/(4\pi)$ in comparison with measured peak frequencies as a function of different applied magnetic field intensities $B_0$. Both curves are in reasonable agreement, which suggests that the superimposed swirl flow can well reflect the dependence of resonance frequencies on $B_0$, although this comparison does not allow to draw an irrefutable conclusion due to the correction factor only determined empirically. On the basis of these findings,
we have always shifted the natural frequencies by the frequency
$f_{\rm swirl}$ for all the comparisons of the wave model against measured resonance curves presented below.  
\subsection{Resonance curves}
\begin{figure}
	\hspace*{-0.5cm}
	\vspace*{0cm}
	\centering
	\includegraphics[width=1.02\columnwidth]{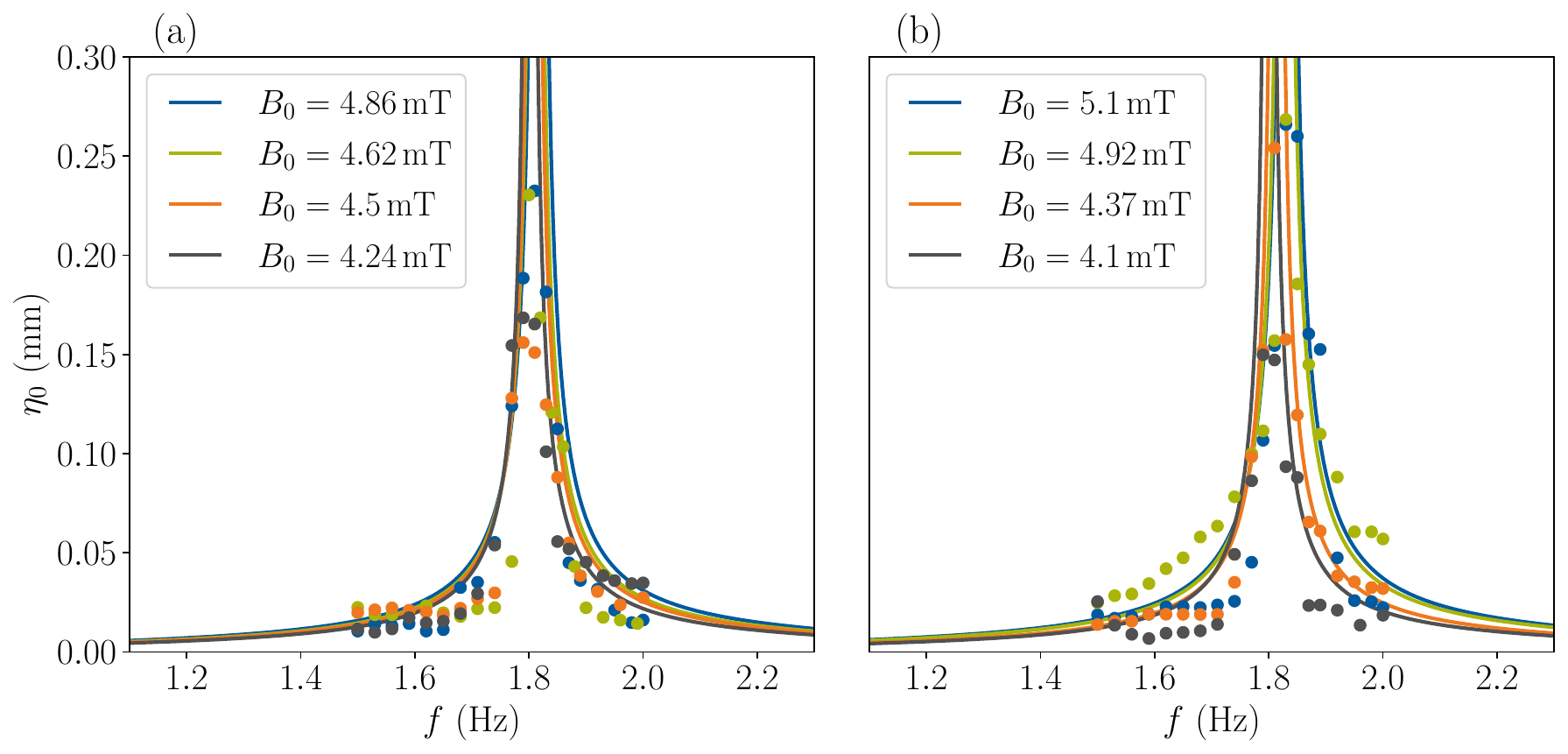}
	\caption{Wave amplitudes $\eta_0$ for different applied RMF intensities $B_0$ as a function of the RMF frequency $f$ measured in $h_1 = 3.5\,{\rm cm}$, $h_2 = 4\,{\rm cm}$ (a) and $h_1 = 3\,{\rm cm}$, $h_2 = 4.5\,{\rm cm}$ (b) stratifications. The dots mark individual measurements and the solid lines represent the predicted resonance curves (Eq.\,(\ref{Eq:ElevNonDim})) with corrected resonance frequencies due to Eq.\,(\ref{Eq:SwirlFreq}). \label{f:ResCurves}}
\end{figure}
We have always only seen a clear peak at the sensor position $r=4.2\,{\rm cm}$ associated with the first resonance $f = \omega_{21}/(4\pi)$, see Fig.\,\ref{f:ExampleCurve}, for which reason we have devoted a number of finer frequency step-resolved measurement campaigns to this region. We noticed that the measured wave amplitudes were subject at times to significant fluctuations in the order of $\delta_{\eta} =  0.03\,{\rm mm}$. Measurements were only reproducible within the limits of fluctuations; repeating the same measurement occasionally yielded slightly different values. We might also have observed small hysteresis effects in some cases. Starting with frequencies lower than the peak frequency $f < \omega_{21}/(4\pi)$ and step-wise increasing or starting with higher frequencies than the peak frequency $f > \omega_{21}/(4\pi)$ and step-wise reducing $f$ sometimes resulted in slightly misplaced resonance profiles. However, the differences tended to be within the range of statistical fluctuations, preventing these effects from being clearly attributed to hysteresis. Despite all these uncertainties, the wave response can be clearly assigned to the irrotational part of the Lorentz force treated in our model. Fig.\,\ref{f:ResCurves} shows different representative resonance curves in comparison with the measurements for differently selected magnetic fields $B_0$ of two different stratification examples. Particularly at lower amplitudes, the measurement points spread considerably around the theoretical curves. In most cases, the theory somewhat overestimates measured amplitudes, in a few cases measurements have also exceeded the predictions. For higher amplitudes $\eta_0 \gtrsim 0.1\,{\rm mm}$, where the interface elevation more closely resembles the sinusoidal signal and wave amplitudes could be determined more precisely, most measurements agree remarkably well with the model predictions. The width of the peak, which is represented quite well, depends very sensitively on $B_0$ and the geometric parameters. It can therefore be concluded that, despite the many intricacies associated with the superimposed swirling flow, the presence of higher harmonics and secondary flow effects such as Ekman pumping, the dominant wave motion can be clearly ascribed to the irrotational part of the Lorentz force modeled in this study.  
\subsection{Peak amplitudes}
\begin{figure}
	\vspace*{0cm}
	\hspace*{-0.5cm}
	\centering
	\includegraphics[width=0.8\columnwidth]{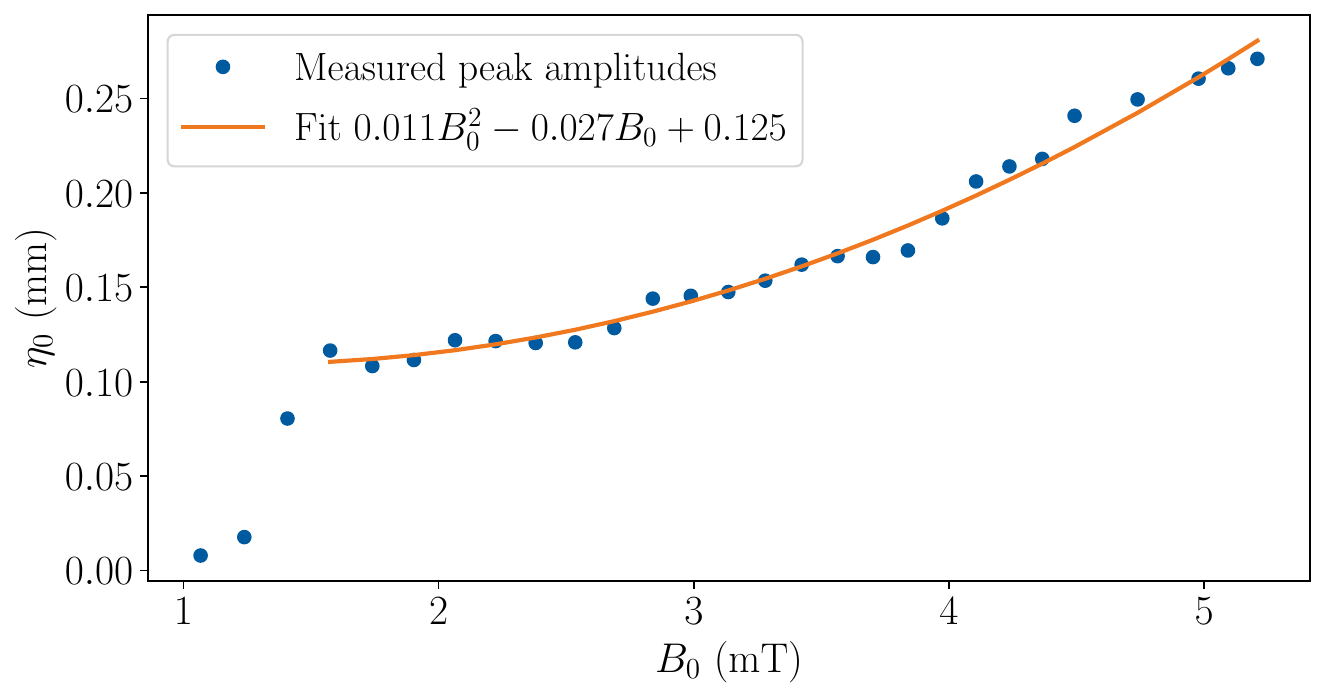}
	\caption{Measured peak amplitudes $\eta_0$ as a function of the applied magnetic field $B_0$. The orange line shows a square fit for all measurements accounting from the fourth value onwards with an coefficient of determination of $R^2 = 0.984$. \label{f:PeakAmps}}
\end{figure}
In a separate measurement campaign with the stratification $h_1 = 3.5\,{\rm cm}$, $h_2 = 4\,{\rm cm}$, we carefully recorded the highest resonant peak amplitudes for different applied $B_0$. This was achieved by varying the frequency in small steps around the resonance frequency and always noting the highest observed amplitude value. Contrary to non-resonant wave responses, peak amplitudes are governed by dissipation mechanisms present in the system. If we subsume all dissipation sources into one damping rate $\lambda$, linear peak amplitudes scale as \cite{Horstmann2020}
\begin{align}
	\frac{\eta_0}{R} \sim Fr \frac{\omega_{21}}{\lambda}
\end{align}
and are therefore predicted to grow quadratically with the magnetic field since $Fr \sim B_0^2$. As can be seen in Fig.\,{\ref{f:PeakAmps}}, we have indeed observed such a quadratic increase for magnetic fields larger than $B_0 = 1.5\,{\rm mT}$. Prior to that, however, the scaling behavior was exceptional. No significant wave amplitudes were measured for smaller values. From $B=1.3\,{\rm mT}$, the amplitude jumps up rapidly and then follows the expected quadratic curve. The fitted quadratic curve thereby contains an offset that is not reflected in our model. It is conceivable that the Lorentz force cannot overcome viscosity for very small $B_0$ and no wave is created. In addition, damping in the system is not solely caused by viscous friction, but also by the magnetic field itself. As the wave rotates faster than the RMF, an electric current is induced, which, according to Lenz's rule, results in a Lorentz force that counteracts its causative motion. This magnetic damping, which itself also depends on $B_0$, usually quadratically, complicates the scaling behavior. Due to these imponderabilities, the observed resonant peak amplitudes cannot be completely explained by our model. 
      
\subsection{Explanation for the higher harmonics}
\label{Sec:Harmonics}
We will complete our study with an explanation for the detected higher harmonics, which usually do not appear in linear wave regimes. In our case of magnetically excited interfacial waves, the driving Lorentz force itself is nonlinear and receives feedback from the rotating wave. In the simplified Lorentz force (\ref{eq:Lorentz}) considered in our model, we have neglected RMF-induced flow velocities $u,v,w \ll r\Omega$. These force components are small compared to the Lorentz force directly caused by RMF, which is why our model can correctly describe non-resonant wave amplitudes associated with the dominant wave mode $(2,1)$. However, these flow-induced force components are apparently strong enough to impair the amplitude signals with higher frequency components, see Fig.\,{\ref{f:Signal}}. The flow velocities are composed of the swirling flow driven by the mean part of the Lorentz force, secondary flows and flows relating to the irrotational wave motion. The latter are responsible for the higher harmonics. In the Lorentz force, we have velocity-dependent terms scaling as
\begin{align}
	f \sim \operatorname{Re}[(u,v,w)\exp(i2\Omega t - i2\theta)].
\end{align}
From the potential solutions (\ref{eq:Pot1}) and (\ref{eq:Pot2}) we know that wave related velocities show the same harmonic dependency
\begin{align}
	(u,v,w) \sim \operatorname{Re}[\exp(i2\Omega t - i2\theta)]],
\end{align}
such that one obtains products of harmonic terms always involving multiples of $2\Omega$
\begin{align}
	\exp(i2\Omega t - i2\theta)\exp(i2\Omega t - i2\theta) = \exp(i4\Omega t - i4\theta).
\end{align} 
This results in an additional $4\Omega$ forcing that accordingly drives subsidiary (wave) flows with twice the fundamental frequency and we end up with a flow velocity encompassing also the second harmonic frequency   
\begin{align}
	(u,v,w) \sim \operatorname{Re}[\exp(i2\Omega t - i2\theta)] + \exp(i4\Omega t - i4\theta)].
\end{align}
The second harmonic again reintegrates into the Lorentz force and yields the third harmonic
\begin{align}
	\exp(i2\Omega t - i2\theta)\exp(i4\Omega t - i4\theta) = \exp(i6\Omega t - i6\theta).
\end{align}
In this way it becomes clear how the complete harmonic series is unfolded recursively through an electromagnetic feedback mechanism. The higher harmonic force components, however, become increasingly marginal, so that at a certain point higher frequencies can no longer be seen in the amplitude signal. In the resonant cases, higher harmonic perturbations up to $8\cdot2\Omega$ were directly recognizable in echo, see Fig.\,{\ref{f:Signal}}(b). Beyond that, harmonics were still be apparent in the Lomb-Scargle spectrum up to $12 \cdot2\Omega$ or $13\cdot2\Omega$. Higher frequencies than those could not be resolved in our experiments due to the Nyquist limit.    
\section{Concluding remarks}
We have demonstrated that beyond the generation of well-known and widely studied swirling flows, rotating magnetic fields (RMF) can also be used to excite considerable wave motions at free-surfaces and interfaces of electrically conducting liquids.
This requires the RMF to operate in the low-frequency regime of about 1 to $10\, {\rm Hz}$ since in this range the oscillatory part of the Lorentz force, which is irrotational in the leading order and drives the waves through pressure variations, cannot be neglected with respect to the mean part of the Lorentz force. Moreover, the resonance conditions of typical magnetic mould stirrers ($R \sim 10\,{\rm cm}$) are met in this range. Our theoretical model has revealed that a homogeneous RMF always excites $m=2$ waves (two crest-trough pairs along the circumference), and that the first and dominant wave mode ($m=2,n=1$) appearing at the resonance condition $\Omega = \omega_{21}/2$ closely resembles the shape of a hyperbolic paraboloid. This is in contrast to surface waves excited by alternating axial magnetic fields that force axisymmetric $m=0$ standing waves, albeit traveling $m\geq 1$ waves may nevertheless arise from a parametric instability \cite{Galpin1992a}. We further found that all wave modes oscillate with twice the frequency $\omega = 2\Omega$ of the applied RMF and resonance always appears at half the natural frequency $\Omega = \omega_{2n}/2$ of the excited wave mode $n$. 

In our wave experiment, the parabolic waveform of the first resonant mode could be precisely reconstructed and verified by utilizing an arrangement of ten ultrasonic transducers, which facilitate precise point by point measurements of local interface elevations. The excellent acoustic reflectance of the KOH-GaInSn interface has made it possible to measure amplitudes with a resolution of the order of $0.01\,{\rm mm}$. Moreover, quantitative comparison of the measured non-resonant wave amplitudes with the theoretical predictions yielded a good agreement within the measurement uncertainties. At the same time the measurement results also clearly demonstrated the limitations of our linearized wave model. We concluded that the resulting wave dynamics must always be understood as a superposition of a swirling flow, driven by the mean part of the Lorentz force, with an irrotational (but rotating) wave motion described by our model. Owing to the wave propagation in the rotating reference system of the swirling flow, the resonance frequencies measured by us in a stationary system of reference appear to be shifted and were always about $10\, \%$ higher than the theoretical resonance frequencies. By estimating the magnitude of the swirling flow, this virtual frequency shift could be eliminated, though. The experiments also revealed that all captured amplitude signals contained several higher harmonics, which were not represented in linear wave theories that always predict single-frequency responses to single-frequency excitations. The Lorentz force induced by the RMF, however, is by itself nonlinear and introduces higher harmonics recursively through an electromagnetic feedback mechanism, even in small-amplitude regimes. 

Together with the obtained results, this could be a promising starting point for future studies to gain a better understanding of the overall dynamics, including the swirl flow, higher harmonic wave responses and secondary flows arising, e.g., from Ekman pumping. The modeling of the latter and the calculation of magnetic dissipation is essential for computing wave damping rates, by which it finally would become possible to estimate resonant wave responses as well. In this study, we only applied very small magnetic fields up to $B_0 = 5\,{\rm mT}$ in order to keep the influence of the swirling flow as small as possible. Industrial mould stirrers typically operate with strong magnetic fields of $B_0 = 100\,{\rm mT}$ and more. For such cases, the wave model already predicts substantial non-resonant wave amplitudes in the order of centimeters. If melts that involve a free surface or interface are stirred in industrial processes, RMF waves can be of high practical importance. In the most unfavorable case, some resonance frequency may be met during operation and the melt's surface could begin to splash out of control. In the best case, however, surface waves could also be exploited for the purpose of enhancing the mass transfer and homogenization of the melt, or also to facilitate the entrapment of floating melt additives.

\begin{acknowledgments}
This study has received funding from the Deutsche
Forschungsgemeinschaft (DFG, German Research Foundation) by award number 512131026 (to G. M. Horstmann), 	
from the Minerva foundation (to Y. Nezihovski) and from the Israel Science Foundation, grant No. 1363/23 (to A. Gelfgat). G. M. Horstmann and Y. Nezihovski contributed equally to this work, G.M. Horstmann was responsible for the theoretical part, and Y. Nezihovski - for the experimental part of the study.
\end{acknowledgments}
\bibliography{Paper}

\begin{thebibliography}{50}%
\makeatletter
\providecommand \@ifxundefined [1]{%
 \@ifx{#1\undefined}
}%
\providecommand \@ifnum [1]{%
 \ifnum #1\expandafter \@firstoftwo
 \else \expandafter \@secondoftwo
 \fi
}%
\providecommand \@ifx [1]{%
 \ifx #1\expandafter \@firstoftwo
 \else \expandafter \@secondoftwo
 \fi
}%
\providecommand \natexlab [1]{#1}%
\providecommand \enquote  [1]{``#1''}%
\providecommand \bibnamefont  [1]{#1}%
\providecommand \bibfnamefont [1]{#1}%
\providecommand \citenamefont [1]{#1}%
\providecommand \href@noop [0]{\@secondoftwo}%
\providecommand \href [0]{\begingroup \@sanitize@url \@href}%
\providecommand \@href[1]{\@@startlink{#1}\@@href}%
\providecommand \@@href[1]{\endgroup#1\@@endlink}%
\providecommand \@sanitize@url [0]{\catcode `\\12\catcode `\$12\catcode
  `\&12\catcode `\#12\catcode `\^12\catcode `\_12\catcode `\%12\relax}%
\providecommand \@@startlink[1]{}%
\providecommand \@@endlink[0]{}%
\providecommand \url  [0]{\begingroup\@sanitize@url \@url }%
\providecommand \@url [1]{\endgroup\@href {#1}{\urlprefix }}%
\providecommand \urlprefix  [0]{URL }%
\providecommand \Eprint [0]{\href }%
\providecommand \doibase [0]{https://doi.org/}%
\providecommand \selectlanguage [0]{\@gobble}%
\providecommand \bibinfo  [0]{\@secondoftwo}%
\providecommand \bibfield  [0]{\@secondoftwo}%
\providecommand \translation [1]{[#1]}%
\providecommand \BibitemOpen [0]{}%
\providecommand \bibitemStop [0]{}%
\providecommand \bibitemNoStop [0]{.\EOS\space}%
\providecommand \EOS [0]{\spacefactor3000\relax}%
\providecommand \BibitemShut  [1]{\csname bibitem#1\endcsname}%
\let\auto@bib@innerbib\@empty
\bibitem [{\citenamefont {Gelfgat}\ and\ \citenamefont
  {Priede}(1995)}]{Gelfgat1995}%
  \BibitemOpen
  \bibfield  {author} {\bibinfo {author} {\bibfnamefont {Y.~M.}\ \bibnamefont
  {Gelfgat}}\ and\ \bibinfo {author} {\bibfnamefont {J.}~\bibnamefont
  {Priede}},\ }\bibfield  {title} {\bibinfo {title} {{{MHD}} flows in a
  rotating magnetic field (a review)},\ }\href@noop {} {\bibfield  {journal}
  {\bibinfo  {journal} {Magnetohydrodynamics}\ }\textbf {\bibinfo {volume}
  {31}},\ \bibinfo {pages} {188} (\bibinfo {year} {1995})}\BibitemShut
  {NoStop}%
\bibitem [{\citenamefont {Davidson}(2001)}]{Davidson2001}%
  \BibitemOpen
  \bibfield  {author} {\bibinfo {author} {\bibfnamefont {P.~A.}\ \bibnamefont
  {Davidson}},\ }\href {https://doi.org/10.1017/CBO9780511626333} {\emph
  {\bibinfo {title} {An {{Introduction}} to {{Magnetohydrodynamics}}}}}\
  (\bibinfo  {publisher} {{Cambridge University Press}},\ \bibinfo {year}
  {2001})\BibitemShut {NoStop}%
\bibitem [{\citenamefont {Spitzer}\ \emph {et~al.}(1986)\citenamefont
  {Spitzer}, \citenamefont {Dubke},\ and\ \citenamefont
  {Schwerdtfeger}}]{Spitzer1986}%
  \BibitemOpen
  \bibfield  {author} {\bibinfo {author} {\bibfnamefont {K.-H.}\ \bibnamefont
  {Spitzer}}, \bibinfo {author} {\bibfnamefont {M.}~\bibnamefont {Dubke}},\
  and\ \bibinfo {author} {\bibfnamefont {K.}~\bibnamefont {Schwerdtfeger}},\
  }\bibfield  {title} {\bibinfo {title} {Rotational electromagnetic stirring in
  continuous casting of round strands},\ }\href
  {https://doi.org/10.1007/BF02670825} {\bibfield  {journal} {\bibinfo
  {journal} {Metall. Trans. B.}\ }\textbf {\bibinfo {volume} {17}},\ \bibinfo
  {pages} {119} (\bibinfo {year} {1986})}\BibitemShut {NoStop}%
\bibitem [{\citenamefont {Kunstreich}(2003)}]{Kunstreich2003}%
  \BibitemOpen
  \bibfield  {author} {\bibinfo {author} {\bibfnamefont {S.}~\bibnamefont
  {Kunstreich}},\ }\bibfield  {title} {\bibinfo {title} {Electromagnetic
  stirring for continuous casting},\ }\href
  {https://doi.org/10.1051/metal:2003198} {\bibfield  {journal} {\bibinfo
  {journal} {Rev. Met. Paris}\ }\textbf {\bibinfo {volume} {100}},\ \bibinfo
  {pages} {395} (\bibinfo {year} {2003})}\BibitemShut {NoStop}%
\bibitem [{\citenamefont {Gelfgat}\ \emph {et~al.}(1999)\citenamefont
  {Gelfgat}, \citenamefont {Krumin},\ and\ \citenamefont
  {Abricka}}]{Gelfgat1999a}%
  \BibitemOpen
  \bibfield  {author} {\bibinfo {author} {\bibfnamefont {{\relax Yu}.~M.}\
  \bibnamefont {Gelfgat}}, \bibinfo {author} {\bibfnamefont {J.}~\bibnamefont
  {Krumin}},\ and\ \bibinfo {author} {\bibfnamefont {M.}~\bibnamefont
  {Abricka}},\ }\bibfield  {title} {\bibinfo {title} {Rotating magnetic fields
  as a means to control the hydrodynamics and heat transfer in single crystal
  growth processes},\ }\href {https://doi.org/10.1016/S0960-8974(99)00009-1}
  {\bibfield  {journal} {\bibinfo  {journal} {Prog. Cryst. Growth Charact.
  Mater.}\ }\textbf {\bibinfo {volume} {38}},\ \bibinfo {pages} {73} (\bibinfo
  {year} {1999})}\BibitemShut {NoStop}%
\bibitem [{\citenamefont {Gelfgat}\ \emph {et~al.}(2001)\citenamefont
  {Gelfgat}, \citenamefont {Abricka},\ and\ \citenamefont {Kr{\=u}mi{\c n}{\v
  s}}}]{Gelfgat2001}%
  \BibitemOpen
  \bibfield  {author} {\bibinfo {author} {\bibfnamefont {{\relax
  Yu}.}~\bibnamefont {Gelfgat}}, \bibinfo {author} {\bibfnamefont
  {M.}~\bibnamefont {Abricka}},\ and\ \bibinfo {author} {\bibfnamefont
  {J.}~\bibnamefont {Kr{\=u}mi{\c n}{\v s}}},\ }\bibfield  {title} {\bibinfo
  {title} {Influence of alternating magnetic field on the hydrodynamics and
  heat/mass transfer in the processes of bulk single crystal growth},\
  }\href@noop {} {\bibfield  {journal} {\bibinfo  {journal}
  {Magnetohydrodynamics}\ }\textbf {\bibinfo {volume} {37}},\ \bibinfo {pages}
  {337} (\bibinfo {year} {2001})}\BibitemShut {NoStop}%
\bibitem [{\citenamefont {Moffatt}(1965)}]{Moffatt1965}%
  \BibitemOpen
  \bibfield  {author} {\bibinfo {author} {\bibfnamefont {H.~K.}\ \bibnamefont
  {Moffatt}},\ }\bibfield  {title} {\bibinfo {title} {On fluid flow induced by
  a rotating magnetic field},\ }\href
  {https://doi.org/10.1017/S0022112065000940} {\bibfield  {journal} {\bibinfo
  {journal} {J. Fluid Mech.}\ }\textbf {\bibinfo {volume} {22}},\ \bibinfo
  {pages} {521} (\bibinfo {year} {1965})}\BibitemShut {NoStop}%
\bibitem [{\citenamefont {Dahlberg}(1972)}]{Dahlberg1972}%
  \BibitemOpen
  \bibfield  {author} {\bibinfo {author} {\bibfnamefont {E.}~\bibnamefont
  {Dahlberg}},\ }\href@noop {} {\emph {\bibinfo {title} {On the Action of a
  Rotating Magnetic Field on a Conducting Liquid}}},\ \bibinfo {type} {Tech.
  Rep.}\ \bibinfo {number} {AE--447}\ (\bibinfo  {institution} {{AB-Atomenergi
  Rep.}},\ \bibinfo {address} {{Sweden}},\ \bibinfo {year} {1972})\BibitemShut
  {NoStop}%
\bibitem [{\citenamefont {Davidson}\ and\ \citenamefont
  {Hunt}(1987)}]{Davidson1987}%
  \BibitemOpen
  \bibfield  {author} {\bibinfo {author} {\bibfnamefont {P.~A.}\ \bibnamefont
  {Davidson}}\ and\ \bibinfo {author} {\bibfnamefont {J.~C.~R.}\ \bibnamefont
  {Hunt}},\ }\bibfield  {title} {\bibinfo {title} {Swirling recirculating flow
  an a liquid-metal column generated by a rotating magnetic field},\ }\href
  {https://doi.org/10.1017/S0022112087003082} {\bibfield  {journal} {\bibinfo
  {journal} {J. Fluid Mech.}\ }\textbf {\bibinfo {volume} {185}},\ \bibinfo
  {pages} {67} (\bibinfo {year} {1987})}\BibitemShut {NoStop}%
\bibitem [{\citenamefont {Sneyd}(1993)}]{Sneyd1993}%
  \BibitemOpen
  \bibfield  {author} {\bibinfo {author} {\bibfnamefont {A.}~\bibnamefont
  {Sneyd}},\ }\bibfield  {title} {\bibinfo {title} {Theory of electromagnetic
  stirring by {{AC}} fields},\ }\href {https://doi.org/10.1093/imaman/5.1.87}
  {\bibfield  {journal} {\bibinfo  {journal} {IMA J. Manag. Math.}\ }\textbf
  {\bibinfo {volume} {5}},\ \bibinfo {pages} {87} (\bibinfo {year}
  {1993})}\BibitemShut {NoStop}%
\bibitem [{\citenamefont {Priede}\ and\ \citenamefont
  {Gelfgat}(1996)}]{Priede1996}%
  \BibitemOpen
  \bibfield  {author} {\bibinfo {author} {\bibfnamefont {J.}~\bibnamefont
  {Priede}}\ and\ \bibinfo {author} {\bibfnamefont {Y.~M.}\ \bibnamefont
  {Gelfgat}},\ }\bibfield  {title} {\bibinfo {title} {Mathematical model of the
  mean electromagnetic force induced by a rotating magnetic field a liquid
  column of a finite length},\ }\href@noop {} {\bibfield  {journal} {\bibinfo
  {journal} {Magnetohydrodynamics}\ }\textbf {\bibinfo {volume} {32}},\
  \bibinfo {pages} {249} (\bibinfo {year} {1996})}\BibitemShut {NoStop}%
\bibitem [{\citenamefont {Witkowski}\ and\ \citenamefont
  {Marty}(1998)}]{Witkowski1998}%
  \BibitemOpen
  \bibfield  {author} {\bibinfo {author} {\bibfnamefont {L.~M.}\ \bibnamefont
  {Witkowski}}\ and\ \bibinfo {author} {\bibfnamefont {P.}~\bibnamefont
  {Marty}},\ }\bibfield  {title} {\bibinfo {title} {Effect of a rotating
  magnetic field of arbitrary frequency on a liquid metal column},\ }\href
  {https://doi.org/10.1016/S0997-7546(98)80061-7} {\bibfield  {journal}
  {\bibinfo  {journal} {Eur. J. Mech. B/Fluids}\ }\textbf {\bibinfo {volume}
  {17}},\ \bibinfo {pages} {239} (\bibinfo {year} {1998})}\BibitemShut
  {NoStop}%
\bibitem [{\citenamefont {Grants}\ and\ \citenamefont
  {Gerbeth}(2001)}]{Grants2001}%
  \BibitemOpen
  \bibfield  {author} {\bibinfo {author} {\bibfnamefont {I.}~\bibnamefont
  {Grants}}\ and\ \bibinfo {author} {\bibfnamefont {G.}~\bibnamefont
  {Gerbeth}},\ }\bibfield  {title} {\bibinfo {title} {Stability of axially
  symmetric flow driven by a rotating magnetic field in a cylindrical cavity},\
  }\href {https://doi.org/10.1017/S0022112000003141} {\bibfield  {journal}
  {\bibinfo  {journal} {J. Fluid Mech.}\ }\textbf {\bibinfo {volume} {431}},\
  \bibinfo {pages} {407} (\bibinfo {year} {2001})}\BibitemShut {NoStop}%
\bibitem [{\citenamefont {Grants}\ and\ \citenamefont
  {Gerbeth}(2002)}]{Grants2002}%
  \BibitemOpen
  \bibfield  {author} {\bibinfo {author} {\bibfnamefont {I.}~\bibnamefont
  {Grants}}\ and\ \bibinfo {author} {\bibfnamefont {G.}~\bibnamefont
  {Gerbeth}},\ }\bibfield  {title} {\bibinfo {title} {Linear three-dimensional
  instability of a magnetically driven rotating flow},\ }\href
  {https://doi.org/10.1017/S0022112002008807} {\bibfield  {journal} {\bibinfo
  {journal} {J. Fluid Mech.}\ }\textbf {\bibinfo {volume} {463}},\ \bibinfo
  {pages} {229} (\bibinfo {year} {2002})}\BibitemShut {NoStop}%
\bibitem [{\citenamefont {Davidson}\ \emph {et~al.}(1995)\citenamefont
  {Davidson}, \citenamefont {Short},\ and\ \citenamefont
  {Kinnear}}]{Davidson1995}%
  \BibitemOpen
  \bibfield  {author} {\bibinfo {author} {\bibfnamefont {P.~A.}\ \bibnamefont
  {Davidson}}, \bibinfo {author} {\bibfnamefont {D.~J.}\ \bibnamefont
  {Short}},\ and\ \bibinfo {author} {\bibfnamefont {D.}~\bibnamefont
  {Kinnear}},\ }\bibfield  {title} {\bibinfo {title} {The {{Role}} of {{Ekman
  Pumping}} in {{Confined}}, {{Electromagnetically-Driven Flows}}},\
  }\href@noop {} {\bibfield  {journal} {\bibinfo  {journal} {Eur. J. Mech.
  B/Fluids}\ }\textbf {\bibinfo {volume} {14}},\ \bibinfo {pages} {795}
  (\bibinfo {year} {1995})}\BibitemShut {NoStop}%
\bibitem [{\citenamefont {Nikrityuk}\ \emph {et~al.}(2004)\citenamefont
  {Nikrityuk}, \citenamefont {Eckert},\ and\ \citenamefont
  {Grundmann}}]{Nikrityuk2004}%
  \BibitemOpen
  \bibfield  {author} {\bibinfo {author} {\bibfnamefont {P.~A.}\ \bibnamefont
  {Nikrityuk}}, \bibinfo {author} {\bibfnamefont {K.}~\bibnamefont {Eckert}},\
  and\ \bibinfo {author} {\bibfnamefont {R.}~\bibnamefont {Grundmann}},\
  }\bibfield  {title} {\bibinfo {title} {Numerical study of spin-up dynamics of
  a liquid metal stirred by rotating magnetic fields in a cylinder with the
  apper free surface},\ }\href@noop {} {\bibfield  {journal} {\bibinfo
  {journal} {Magnetohydrodynamics}\ }\textbf {\bibinfo {volume} {40}},\
  \bibinfo {pages} {127} (\bibinfo {year} {2004})}\BibitemShut {NoStop}%
\bibitem [{\citenamefont {Nikrityuk}\ \emph {et~al.}(2005)\citenamefont
  {Nikrityuk}, \citenamefont {Ungarish}, \citenamefont {Eckert},\ and\
  \citenamefont {Grundmann}}]{Nikrityuk2005}%
  \BibitemOpen
  \bibfield  {author} {\bibinfo {author} {\bibfnamefont {P.~A.}\ \bibnamefont
  {Nikrityuk}}, \bibinfo {author} {\bibfnamefont {M.}~\bibnamefont {Ungarish}},
  \bibinfo {author} {\bibfnamefont {K.}~\bibnamefont {Eckert}},\ and\ \bibinfo
  {author} {\bibfnamefont {R.}~\bibnamefont {Grundmann}},\ }\bibfield  {title}
  {\bibinfo {title} {Spin-up of a liquid metal flow driven by a rotating
  magnetic field in a finite cylinder: {{A}} numerical and an analytical
  study},\ }\href {https://doi.org/10.1063/1.1897323} {\bibfield  {journal}
  {\bibinfo  {journal} {Phys. Fluids}\ }\textbf {\bibinfo {volume} {17}},\
  \bibinfo {pages} {067101} (\bibinfo {year} {2005})}\BibitemShut {NoStop}%
\bibitem [{\citenamefont {R{\"a}biger}\ \emph {et~al.}(2010)\citenamefont
  {R{\"a}biger}, \citenamefont {Eckert},\ and\ \citenamefont
  {Gerbeth}}]{Rabiger2010}%
  \BibitemOpen
  \bibfield  {author} {\bibinfo {author} {\bibfnamefont {D.}~\bibnamefont
  {R{\"a}biger}}, \bibinfo {author} {\bibfnamefont {S.}~\bibnamefont
  {Eckert}},\ and\ \bibinfo {author} {\bibfnamefont {G.}~\bibnamefont
  {Gerbeth}},\ }\bibfield  {title} {\bibinfo {title} {Measurements of an
  unsteady liquid metal flow during spin-up driven by a rotating magnetic
  field},\ }\href {https://doi.org/10.1007/s00348-009-0735-1} {\bibfield
  {journal} {\bibinfo  {journal} {Exp. Fluids}\ }\textbf {\bibinfo {volume}
  {48}},\ \bibinfo {pages} {233} (\bibinfo {year} {2010})}\BibitemShut
  {NoStop}%
\bibitem [{\citenamefont {Vogt}\ \emph {et~al.}(2012)\citenamefont {Vogt},
  \citenamefont {Grants}, \citenamefont {R{\"a}biger}, \citenamefont {Eckert},\
  and\ \citenamefont {Gerbeth}}]{Vogt2012}%
  \BibitemOpen
  \bibfield  {author} {\bibinfo {author} {\bibfnamefont {T.}~\bibnamefont
  {Vogt}}, \bibinfo {author} {\bibfnamefont {I.}~\bibnamefont {Grants}},
  \bibinfo {author} {\bibfnamefont {D.}~\bibnamefont {R{\"a}biger}}, \bibinfo
  {author} {\bibfnamefont {S.}~\bibnamefont {Eckert}},\ and\ \bibinfo {author}
  {\bibfnamefont {G.}~\bibnamefont {Gerbeth}},\ }\bibfield  {title} {\bibinfo
  {title} {On the formation of {{Taylor}}{\textendash}{{G{\"o}rtler}} vortices
  in {{RMF-driven}} spin-up flows},\ }\href
  {https://doi.org/10.1007/s00348-011-1196-x} {\bibfield  {journal} {\bibinfo
  {journal} {Exp. Fluids}\ }\textbf {\bibinfo {volume} {52}},\ \bibinfo {pages}
  {1} (\bibinfo {year} {2012})}\BibitemShut {NoStop}%
\bibitem [{\citenamefont {Travnikov}\ \emph {et~al.}(2012)\citenamefont
  {Travnikov}, \citenamefont {Eckert}, \citenamefont {Nikrityuk}, \citenamefont
  {Odenbach}, \citenamefont {Vogt},\ and\ \citenamefont
  {Eckert}}]{Travnikov2012}%
  \BibitemOpen
  \bibfield  {author} {\bibinfo {author} {\bibfnamefont {V.}~\bibnamefont
  {Travnikov}}, \bibinfo {author} {\bibfnamefont {K.}~\bibnamefont {Eckert}},
  \bibinfo {author} {\bibfnamefont {P.~A.}\ \bibnamefont {Nikrityuk}}, \bibinfo
  {author} {\bibfnamefont {S.}~\bibnamefont {Odenbach}}, \bibinfo {author}
  {\bibfnamefont {T.}~\bibnamefont {Vogt}},\ and\ \bibinfo {author}
  {\bibfnamefont {S.}~\bibnamefont {Eckert}},\ }\bibfield  {title} {\bibinfo
  {title} {Flow oscillations driven by a rotating magnetic field in liquid
  metal columns with an upper free surface},\ }\href
  {https://doi.org/10.1016/j.jcrysgro.2011.10.048} {\bibfield  {journal}
  {\bibinfo  {journal} {J. Cryst. Growth}\ }\textbf {\bibinfo {volume} {339}},\
  \bibinfo {pages} {52} (\bibinfo {year} {2012})}\BibitemShut {NoStop}%
\bibitem [{\citenamefont {Grants}\ \emph {et~al.}(2008)\citenamefont {Grants},
  \citenamefont {Zhang}, \citenamefont {Eckert},\ and\ \citenamefont
  {Gerbeth}}]{Grants2008}%
  \BibitemOpen
  \bibfield  {author} {\bibinfo {author} {\bibfnamefont {I.}~\bibnamefont
  {Grants}}, \bibinfo {author} {\bibfnamefont {C.}~\bibnamefont {Zhang}},
  \bibinfo {author} {\bibfnamefont {S.}~\bibnamefont {Eckert}},\ and\ \bibinfo
  {author} {\bibfnamefont {G.}~\bibnamefont {Gerbeth}},\ }\bibfield  {title}
  {\bibinfo {title} {Experimental observation of swirl accumulation in a
  magnetically driven flow},\ }\href
  {https://doi.org/10.1017/S0022112008003650} {\bibfield  {journal} {\bibinfo
  {journal} {J. Fluid Mech.}\ }\textbf {\bibinfo {volume} {616}},\ \bibinfo
  {pages} {135} (\bibinfo {year} {2008})}\BibitemShut {NoStop}%
\bibitem [{\citenamefont {Vogt}\ \emph {et~al.}(2013)\citenamefont {Vogt},
  \citenamefont {Grants}, \citenamefont {Eckert},\ and\ \citenamefont
  {Gerbeth}}]{Vogt2013}%
  \BibitemOpen
  \bibfield  {author} {\bibinfo {author} {\bibfnamefont {T.}~\bibnamefont
  {Vogt}}, \bibinfo {author} {\bibfnamefont {I.}~\bibnamefont {Grants}},
  \bibinfo {author} {\bibfnamefont {S.}~\bibnamefont {Eckert}},\ and\ \bibinfo
  {author} {\bibfnamefont {G.}~\bibnamefont {Gerbeth}},\ }\bibfield  {title}
  {\bibinfo {title} {Spin-up of a magnetically driven tornado-like vortex},\
  }\href {https://doi.org/10.1017/jfm.2013.552} {\bibfield  {journal} {\bibinfo
   {journal} {J. Fluid Mech.}\ }\textbf {\bibinfo {volume} {736}},\ \bibinfo
  {pages} {641} (\bibinfo {year} {2013})}\BibitemShut {NoStop}%
\bibitem [{\citenamefont {Grants}\ \emph {et~al.}(2015)\citenamefont {Grants},
  \citenamefont {R{\"a}biger}, \citenamefont {Vogt}, \citenamefont {Eckert},\
  and\ \citenamefont {Gerbeth}}]{Grants2015}%
  \BibitemOpen
  \bibfield  {author} {\bibinfo {author} {\bibfnamefont {I.}~\bibnamefont
  {Grants}}, \bibinfo {author} {\bibfnamefont {D.}~\bibnamefont {R{\"a}biger}},
  \bibinfo {author} {\bibfnamefont {T.}~\bibnamefont {Vogt}}, \bibinfo {author}
  {\bibfnamefont {S.}~\bibnamefont {Eckert}},\ and\ \bibinfo {author}
  {\bibfnamefont {G.}~\bibnamefont {Gerbeth}},\ }\bibfield  {title} {\bibinfo
  {title} {Application of magnetically driven tornado-like vortex for stirring
  floating particles into liquid metal},\ }\href@noop {} {\bibfield  {journal}
  {\bibinfo  {journal} {Magnetohydrodynamics}\ }\textbf {\bibinfo {volume}
  {51}},\ \bibinfo {pages} {419} (\bibinfo {year} {2015})}\BibitemShut
  {NoStop}%
\bibitem [{\citenamefont {Gelfgat}\ \emph {et~al.}(2005)\citenamefont
  {Gelfgat}, \citenamefont {Skopis},\ and\ \citenamefont
  {Grabis}}]{Gelfgat2005}%
  \BibitemOpen
  \bibfield  {author} {\bibinfo {author} {\bibfnamefont {{\relax
  Yu}.}~\bibnamefont {Gelfgat}}, \bibinfo {author} {\bibfnamefont
  {M.}~\bibnamefont {Skopis}},\ and\ \bibinfo {author} {\bibfnamefont
  {J.}~\bibnamefont {Grabis}},\ }\bibfield  {title} {\bibinfo {title}
  {Electromagnetically driven vortex flow to introduce small solid particles
  into liquid metal},\ }\href@noop {} {\bibfield  {journal} {\bibinfo
  {journal} {Magnetohydrodynamics}\ }\textbf {\bibinfo {volume} {41}},\
  \bibinfo {pages} {249} (\bibinfo {year} {2005})}\BibitemShut {NoStop}%
\bibitem [{\citenamefont {Willers}\ \emph {et~al.}(2017)\citenamefont
  {Willers}, \citenamefont {Barna}, \citenamefont {Reiter},\ and\ \citenamefont
  {Eckert}}]{Willers2017}%
  \BibitemOpen
  \bibfield  {author} {\bibinfo {author} {\bibfnamefont {B.}~\bibnamefont
  {Willers}}, \bibinfo {author} {\bibfnamefont {M.}~\bibnamefont {Barna}},
  \bibinfo {author} {\bibfnamefont {J.}~\bibnamefont {Reiter}},\ and\ \bibinfo
  {author} {\bibfnamefont {S.}~\bibnamefont {Eckert}},\ }\bibfield  {title}
  {\bibinfo {title} {Experimental {{Investigations}} of {{Rotary
  Electromagnetic Mould Stirring}} in {{Continuous Casting Using}} a {{Cold
  Liquid Metal Model}}},\ }\href
  {https://doi.org/10.2355/isijinternational.ISIJINT-2016-495} {\bibfield
  {journal} {\bibinfo  {journal} {ISIJ Int.}\ }\textbf {\bibinfo {volume}
  {57}},\ \bibinfo {pages} {468} (\bibinfo {year} {2017})}\BibitemShut
  {NoStop}%
\bibitem [{\citenamefont {Galpin}\ and\ \citenamefont
  {Fautrelle}(1992)}]{Galpin1992}%
  \BibitemOpen
  \bibfield  {author} {\bibinfo {author} {\bibfnamefont {J.~M.}\ \bibnamefont
  {Galpin}}\ and\ \bibinfo {author} {\bibfnamefont {Y.}~\bibnamefont
  {Fautrelle}},\ }\bibfield  {title} {\bibinfo {title} {Liquid-metal flows
  induced by low-frequency alternating magnetic fields},\ }\href
  {https://doi.org/10.1017/S0022112092004452} {\bibfield  {journal} {\bibinfo
  {journal} {J. Fluid Mech.}\ }\textbf {\bibinfo {volume} {239}},\ \bibinfo
  {pages} {383} (\bibinfo {year} {1992})}\BibitemShut {NoStop}%
\bibitem [{\citenamefont {Galpin}\ \emph {et~al.}(1992)\citenamefont {Galpin},
  \citenamefont {Fautrelle},\ and\ \citenamefont {Sneyd}}]{Galpin1992a}%
  \BibitemOpen
  \bibfield  {author} {\bibinfo {author} {\bibfnamefont {J.~M.}\ \bibnamefont
  {Galpin}}, \bibinfo {author} {\bibfnamefont {Y.}~\bibnamefont {Fautrelle}},\
  and\ \bibinfo {author} {\bibfnamefont {A.~D.}\ \bibnamefont {Sneyd}},\
  }\bibfield  {title} {\bibinfo {title} {Parametric resonance in low-frequency
  magnetic stirring},\ }\href {https://doi.org/10.1017/S0022112092004464}
  {\bibfield  {journal} {\bibinfo  {journal} {J. Fluid Mech.}\ }\textbf
  {\bibinfo {volume} {239}},\ \bibinfo {pages} {409} (\bibinfo {year}
  {1992})}\BibitemShut {NoStop}%
\bibitem [{\citenamefont {Fautrelle}\ and\ \citenamefont
  {Sneyd}(2005)}]{Fautrelle2005a}%
  \BibitemOpen
  \bibfield  {author} {\bibinfo {author} {\bibfnamefont {Y.}~\bibnamefont
  {Fautrelle}}\ and\ \bibinfo {author} {\bibfnamefont {A.~D.}\ \bibnamefont
  {Sneyd}},\ }\bibfield  {title} {\bibinfo {title} {Surface waves created by
  low-frequency magnetic fields},\ }\href
  {https://doi.org/10.1016/j.euromechflu.2004.05.005} {\bibfield  {journal}
  {\bibinfo  {journal} {Eur. J. Mech. B/Fluids}\ }\textbf {\bibinfo {volume}
  {24}},\ \bibinfo {pages} {91} (\bibinfo {year} {2005})}\BibitemShut {NoStop}%
\bibitem [{\citenamefont {Fautrelle}\ \emph {et~al.}(2007)\citenamefont
  {Fautrelle}, \citenamefont {Sneyd},\ and\ \citenamefont
  {Etay}}]{Fautrelle2007}%
  \BibitemOpen
  \bibfield  {author} {\bibinfo {author} {\bibfnamefont {Y.}~\bibnamefont
  {Fautrelle}}, \bibinfo {author} {\bibfnamefont {A.}~\bibnamefont {Sneyd}},\
  and\ \bibinfo {author} {\bibfnamefont {J.}~\bibnamefont {Etay}},\ }\bibinfo
  {title} {Effect of {{AC Magnetic Fields}} on {{Free Surfaces}}},\ in\ \href
  {https://doi.org/10.1007/978-1-4020-4833-3_21} {\emph {\bibinfo {booktitle}
  {Magnetohydrodynamics}}},\ Vol.~\bibinfo {volume} {80}\ (\bibinfo
  {publisher} {{Springer Netherlands}},\ \bibinfo {address} {{Dordrecht}},\
  \bibinfo {year} {2007})\ pp.\ \bibinfo {pages} {345--355}\BibitemShut
  {NoStop}%
\bibitem [{\citenamefont {Deng}\ \emph {et~al.}(2011)\citenamefont {Deng},
  \citenamefont {Wang}, \citenamefont {Xu}, \citenamefont {Zhang},\ and\
  \citenamefont {He}}]{Deng2011}%
  \BibitemOpen
  \bibfield  {author} {\bibinfo {author} {\bibfnamefont {A.-y.}\ \bibnamefont
  {Deng}}, \bibinfo {author} {\bibfnamefont {E.-g.}\ \bibnamefont {Wang}},
  \bibinfo {author} {\bibfnamefont {Y.-y.}\ \bibnamefont {Xu}}, \bibinfo
  {author} {\bibfnamefont {X.-w.}\ \bibnamefont {Zhang}},\ and\ \bibinfo
  {author} {\bibfnamefont {J.-c.}\ \bibnamefont {He}},\ }\bibfield  {title}
  {\bibinfo {title} {Oscillation {{Characteristics}} of {{Molten Metal Free
  Surface Under Compound Magnetic Field}}},\ }\href
  {https://doi.org/10.1016/S1006-706X(11)60060-5} {\bibfield  {journal}
  {\bibinfo  {journal} {J. Iron Steel Res. Int.}\ }\textbf {\bibinfo {volume}
  {18}},\ \bibinfo {pages} {25} (\bibinfo {year} {2011})}\BibitemShut {NoStop}%
\bibitem [{\citenamefont {Wu}\ \emph {et~al.}(2020)\citenamefont {Wu},
  \citenamefont {Liu}, \citenamefont {Gao}, \citenamefont {Luo},\ and\
  \citenamefont {Geng}}]{Wu2020}%
  \BibitemOpen
  \bibfield  {author} {\bibinfo {author} {\bibfnamefont {X.}~\bibnamefont
  {Wu}}, \bibinfo {author} {\bibfnamefont {R.}~\bibnamefont {Liu}}, \bibinfo
  {author} {\bibfnamefont {J.}~\bibnamefont {Gao}}, \bibinfo {author}
  {\bibfnamefont {J.}~\bibnamefont {Luo}},\ and\ \bibinfo {author}
  {\bibfnamefont {H.}~\bibnamefont {Geng}},\ }\bibfield  {title} {\bibinfo
  {title} {Analysis on the characteristics of pulsed electromagnetic force and
  the fluctuation behavior of molten metal free surface under pulsed magnetic
  field},\ }\href {https://doi.org/10.1088/2053-1591/ab8df2} {\bibfield
  {journal} {\bibinfo  {journal} {Mater. Res. Express}\ }\textbf {\bibinfo
  {volume} {7}},\ \bibinfo {pages} {056514} (\bibinfo {year}
  {2020})}\BibitemShut {NoStop}%
\bibitem [{\citenamefont {Milgr{\=a}vis}\ \emph {et~al.}(2023)\citenamefont
  {Milgr{\=a}vis}, \citenamefont {Krasti{\c n}{\v s}}, \citenamefont {Kaldre},
  \citenamefont {Kalv{\=a}ns}, \citenamefont {Bojarevi{\v c}s},\ and\
  \citenamefont {Beinerts}}]{Milgravis2023}%
  \BibitemOpen
  \bibfield  {author} {\bibinfo {author} {\bibfnamefont {M.}~\bibnamefont
  {Milgr{\=a}vis}}, \bibinfo {author} {\bibfnamefont {I.}~\bibnamefont
  {Krasti{\c n}{\v s}}}, \bibinfo {author} {\bibfnamefont {I.}~\bibnamefont
  {Kaldre}}, \bibinfo {author} {\bibfnamefont {M.}~\bibnamefont {Kalv{\=a}ns}},
  \bibinfo {author} {\bibfnamefont {A.}~\bibnamefont {Bojarevi{\v c}s}},\ and\
  \bibinfo {author} {\bibfnamefont {T.}~\bibnamefont {Beinerts}},\ }\bibfield
  {title} {\bibinfo {title} {Pulsed and {{Static Magnetic Field Influence}} on
  {{Metallic Alloys}} during {{Solidification}}},\ }\href
  {https://doi.org/10.3390/cryst13020259} {\bibfield  {journal} {\bibinfo
  {journal} {Crystals}\ }\textbf {\bibinfo {volume} {13}},\ \bibinfo {pages}
  {259} (\bibinfo {year} {2023})}\BibitemShut {NoStop}%
\bibitem [{\citenamefont {Debray}\ \emph {et~al.}(1996)\citenamefont {Debray},
  \citenamefont {Fautrelle}, \citenamefont {Burty},\ and\ \citenamefont
  {Galpin}}]{Debray1996}%
  \BibitemOpen
  \bibfield  {author} {\bibinfo {author} {\bibfnamefont {F.}~\bibnamefont
  {Debray}}, \bibinfo {author} {\bibfnamefont {Y.}~\bibnamefont {Fautrelle}},
  \bibinfo {author} {\bibfnamefont {M.}~\bibnamefont {Burty}},\ and\ \bibinfo
  {author} {\bibfnamefont {J.~M.}\ \bibnamefont {Galpin}},\ }\bibfield  {title}
  {\bibinfo {title} {Surface waves and mass transfer induced by a low-frequency
  electromagnetic field},\ }\href@noop {} {\bibfield  {journal} {\bibinfo
  {journal} {Magnetohydrodynamics}\ }\textbf {\bibinfo {volume} {32}},\
  \bibinfo {pages} {122} (\bibinfo {year} {1996})}\BibitemShut {NoStop}%
\bibitem [{\citenamefont {Milgr{\=a}vis}\ \emph {et~al.}(2020)\citenamefont
  {Milgr{\=a}vis}, \citenamefont {Bojarevi{\v c}s}, \citenamefont {Gaile},\
  and\ \citenamefont {Ge{\v z}a}}]{Milgravis2020}%
  \BibitemOpen
  \bibfield  {author} {\bibinfo {author} {\bibfnamefont {M.}~\bibnamefont
  {Milgr{\=a}vis}}, \bibinfo {author} {\bibfnamefont {A.}~\bibnamefont
  {Bojarevi{\v c}s}}, \bibinfo {author} {\bibfnamefont {A.}~\bibnamefont
  {Gaile}},\ and\ \bibinfo {author} {\bibfnamefont {V.}~\bibnamefont {Ge{\v
  z}a}},\ }\bibfield  {title} {\bibinfo {title} {Application of {{AC}} and
  {{DC}} magnetic field for surface wave excitation to enhance mass transfer},\
  }\href {https://doi.org/10.1016/j.jcrysgro.2019.125409} {\bibfield  {journal}
  {\bibinfo  {journal} {J. Cryst. Growth}\ }\textbf {\bibinfo {volume} {534}},\
  \bibinfo {pages} {125409} (\bibinfo {year} {2020})}\BibitemShut {NoStop}%
\bibitem [{\citenamefont {Wiederhold}(2019)}]{Wiederhold2019}%
  \BibitemOpen
  \bibfield  {author} {\bibinfo {author} {\bibfnamefont {A.}~\bibnamefont
  {Wiederhold}},\ }\emph {\bibinfo {title} {{Str{\"o}mungsmessung und
  Str{\"o}mungsbeeinflussung in leitf{\"a}higen Mehrphasensystemen durch
  elektromagnetische Kr{\"a}fte}}},\ \href@noop {} {Ph.D. thesis},\ \bibinfo
  {school} {Technischen Universit{\"a}t Ilmenau}, \bibinfo {address} {{Ilmenau,
  Germany}} (\bibinfo {year} {2019})\BibitemShut {NoStop}%
\bibitem [{\citenamefont {Tagawa}\ and\ \citenamefont
  {Song}(2019)}]{Tagawa2019}%
  \BibitemOpen
  \bibfield  {author} {\bibinfo {author} {\bibfnamefont {T.}~\bibnamefont
  {Tagawa}}\ and\ \bibinfo {author} {\bibfnamefont {K.}~\bibnamefont {Song}},\
  }\bibfield  {title} {\bibinfo {title} {Stability of an {{Axisymmetric Liquid
  Metal Flow Driven}} by a {{Multi-Pole Rotating Magnetic Field}}},\ }\href
  {https://doi.org/10.3390/fluids4020077} {\bibfield  {journal} {\bibinfo
  {journal} {Fluids}\ }\textbf {\bibinfo {volume} {4}},\ \bibinfo {pages} {77}
  (\bibinfo {year} {2019})}\BibitemShut {NoStop}%
\bibitem [{\citenamefont {Horstmann}\ \emph {et~al.}(2020)\citenamefont
  {Horstmann}, \citenamefont {Herreman},\ and\ \citenamefont
  {Weier}}]{Horstmann2020}%
  \BibitemOpen
  \bibfield  {author} {\bibinfo {author} {\bibfnamefont {G.~M.}\ \bibnamefont
  {Horstmann}}, \bibinfo {author} {\bibfnamefont {W.}~\bibnamefont
  {Herreman}},\ and\ \bibinfo {author} {\bibfnamefont {T.}~\bibnamefont
  {Weier}},\ }\bibfield  {title} {\bibinfo {title} {Linear damped interfacial
  wave theory for an orbitally shaken upright circular cylinder},\ }\href
  {https://doi.org/10.1017/jfm.2020.163} {\bibfield  {journal} {\bibinfo
  {journal} {J. Fluid Mech.}\ }\textbf {\bibinfo {volume} {891}},\ \bibinfo
  {pages} {A22} (\bibinfo {year} {2020})}\BibitemShut {NoStop}%
\bibitem [{\citenamefont {Horstmann}(2021)}]{Horstmann2021a}%
  \BibitemOpen
  \bibfield  {author} {\bibinfo {author} {\bibfnamefont {G.~M.}\ \bibnamefont
  {Horstmann}},\ }\emph {\bibinfo {title} {Multilayer Interfacial Wave Dynamics
  in Upright Circular Cylinders with Application to Liquid Metal Batteries}},\
  \href@noop {} {Ph.D. thesis},\ \bibinfo  {school} {Technische Universit{\"a}t
  Dresden}, \bibinfo {address} {{Dresden}} (\bibinfo {year} {2021})\BibitemShut
  {NoStop}%
\bibitem [{\citenamefont {Reclari}\ \emph {et~al.}(2014)\citenamefont
  {Reclari}, \citenamefont {Dreyer}, \citenamefont {Tissot}, \citenamefont
  {Obreschkow}, \citenamefont {Wurm},\ and\ \citenamefont
  {Farhat}}]{Reclari2014}%
  \BibitemOpen
  \bibfield  {author} {\bibinfo {author} {\bibfnamefont {M.}~\bibnamefont
  {Reclari}}, \bibinfo {author} {\bibfnamefont {M.}~\bibnamefont {Dreyer}},
  \bibinfo {author} {\bibfnamefont {S.}~\bibnamefont {Tissot}}, \bibinfo
  {author} {\bibfnamefont {D.}~\bibnamefont {Obreschkow}}, \bibinfo {author}
  {\bibfnamefont {F.~M.}\ \bibnamefont {Wurm}},\ and\ \bibinfo {author}
  {\bibfnamefont {M.}~\bibnamefont {Farhat}},\ }\bibfield  {title} {\bibinfo
  {title} {Surface wave dynamics in orbital shaken cylindrical containers},\
  }\href {https://doi.org/10.1063/1.4874612} {\bibfield  {journal} {\bibinfo
  {journal} {Phys. Fluids}\ }\textbf {\bibinfo {volume} {26}},\ \bibinfo
  {pages} {052104} (\bibinfo {year} {2014})}\BibitemShut {NoStop}%
\bibitem [{\citenamefont {Bongarzone}\ \emph {et~al.}(2022)\citenamefont
  {Bongarzone}, \citenamefont {Guido},\ and\ \citenamefont
  {Gallaire}}]{Bongarzone2022}%
  \BibitemOpen
  \bibfield  {author} {\bibinfo {author} {\bibfnamefont {A.}~\bibnamefont
  {Bongarzone}}, \bibinfo {author} {\bibfnamefont {M.}~\bibnamefont {Guido}},\
  and\ \bibinfo {author} {\bibfnamefont {F.}~\bibnamefont {Gallaire}},\
  }\bibfield  {title} {\bibinfo {title} {An amplitude equation modelling the
  double-crest swirling in orbital-shaken cylindrical containers},\ }\href
  {https://doi.org/10.1017/jfm.2022.440} {\bibfield  {journal} {\bibinfo
  {journal} {J. Fluid Mech.}\ }\textbf {\bibinfo {volume} {943}},\ \bibinfo
  {pages} {A28} (\bibinfo {year} {2022})}\BibitemShut {NoStop}%
\bibitem [{\citenamefont {Case}\ and\ \citenamefont
  {Parkinson}(1957)}]{Case1957}%
  \BibitemOpen
  \bibfield  {author} {\bibinfo {author} {\bibfnamefont {K.~M.}\ \bibnamefont
  {Case}}\ and\ \bibinfo {author} {\bibfnamefont {W.~C.}\ \bibnamefont
  {Parkinson}},\ }\bibfield  {title} {\bibinfo {title} {Damping of surface
  waves in an incompressible liquid},\ }\href
  {https://doi.org/10.1017/S0022112057000051} {\bibfield  {journal} {\bibinfo
  {journal} {J. Fluid Mech.}\ }\textbf {\bibinfo {volume} {2}},\ \bibinfo
  {pages} {172} (\bibinfo {year} {1957})}\BibitemShut {NoStop}%
\bibitem [{\citenamefont {Herreman}\ \emph {et~al.}(2019)\citenamefont
  {Herreman}, \citenamefont {Nore}, \citenamefont {Guermond}, \citenamefont
  {Cappanera}, \citenamefont {Weber},\ and\ \citenamefont
  {Horstmann}}]{Herreman2019a}%
  \BibitemOpen
  \bibfield  {author} {\bibinfo {author} {\bibfnamefont {W.}~\bibnamefont
  {Herreman}}, \bibinfo {author} {\bibfnamefont {C.}~\bibnamefont {Nore}},
  \bibinfo {author} {\bibfnamefont {J.-L.}\ \bibnamefont {Guermond}}, \bibinfo
  {author} {\bibfnamefont {L.}~\bibnamefont {Cappanera}}, \bibinfo {author}
  {\bibfnamefont {N.}~\bibnamefont {Weber}},\ and\ \bibinfo {author}
  {\bibfnamefont {G.~M.}\ \bibnamefont {Horstmann}},\ }\bibfield  {title}
  {\bibinfo {title} {Perturbation theory for metal pad roll instability in
  cylindrical reduction cells},\ }\href {https://doi.org/10.1017/jfm.2019.642}
  {\bibfield  {journal} {\bibinfo  {journal} {J. Fluid Mech.}\ }\textbf
  {\bibinfo {volume} {878}},\ \bibinfo {pages} {598} (\bibinfo {year}
  {2019})}\BibitemShut {NoStop}%
\bibitem [{\citenamefont {Davidson}(1992)}]{Davidson1992}%
  \BibitemOpen
  \bibfield  {author} {\bibinfo {author} {\bibfnamefont {P.~A.}\ \bibnamefont
  {Davidson}},\ }\bibfield  {title} {\bibinfo {title} {Swirling flow in an
  axisymmetric cavity of arbitrary profile, driven by a rotating magnetic
  field},\ }\href {https://doi.org/10.1017/S0022112092000624} {\bibfield
  {journal} {\bibinfo  {journal} {J. Fluid Mech.}\ }\textbf {\bibinfo {volume}
  {245}},\ \bibinfo {pages} {669} (\bibinfo {year} {1992})}\BibitemShut
  {NoStop}%
\bibitem [{\citenamefont {Bouvard}\ \emph {et~al.}(2017)\citenamefont
  {Bouvard}, \citenamefont {Herreman},\ and\ \citenamefont
  {Moisy}}]{Bouvard2017}%
  \BibitemOpen
  \bibfield  {author} {\bibinfo {author} {\bibfnamefont {J.}~\bibnamefont
  {Bouvard}}, \bibinfo {author} {\bibfnamefont {W.}~\bibnamefont {Herreman}},\
  and\ \bibinfo {author} {\bibfnamefont {F.}~\bibnamefont {Moisy}},\ }\bibfield
   {title} {\bibinfo {title} {Mean mass transport in an orbitally shaken
  cylindrical container},\ }\href
  {https://doi.org/10.1103/PhysRevFluids.2.084801} {\bibfield  {journal}
  {\bibinfo  {journal} {Phys. Rev. Fluids}\ }\textbf {\bibinfo {volume} {2}},\
  \bibinfo {pages} {084801} (\bibinfo {year} {2017})}\BibitemShut {NoStop}%
\bibitem [{\citenamefont {Zhang}\ \emph {et~al.}(2011)\citenamefont {Zhang},
  \citenamefont {Shatrov}, \citenamefont {Priede}, \citenamefont {Eckert},\
  and\ \citenamefont {Gerbeth}}]{Zhang2011}%
  \BibitemOpen
  \bibfield  {author} {\bibinfo {author} {\bibfnamefont {C.}~\bibnamefont
  {Zhang}}, \bibinfo {author} {\bibfnamefont {V.}~\bibnamefont {Shatrov}},
  \bibinfo {author} {\bibfnamefont {J.}~\bibnamefont {Priede}}, \bibinfo
  {author} {\bibfnamefont {S.}~\bibnamefont {Eckert}},\ and\ \bibinfo {author}
  {\bibfnamefont {G.}~\bibnamefont {Gerbeth}},\ }\bibfield  {title} {\bibinfo
  {title} {Intermittent {{Behavior Caused}} by {{Surface Oxidation}} in a
  {{Liquid Metal Flow Driven}} by a {{Rotating Magnetic Field}}},\ }\href
  {https://doi.org/10.1007/s11663-011-9538-x} {\bibfield  {journal} {\bibinfo
  {journal} {Metall. Mater. Trans. B.}\ }\textbf {\bibinfo {volume} {42}},\
  \bibinfo {pages} {1188} (\bibinfo {year} {2011})}\BibitemShut {NoStop}%
\bibitem [{\citenamefont {Vogt}\ \emph {et~al.}(2015)\citenamefont {Vogt},
  \citenamefont {Boden}, \citenamefont {Andruszkiewicz}, \citenamefont
  {Eckert}, \citenamefont {Eckert},\ and\ \citenamefont {Gerbeth}}]{Vogt2015}%
  \BibitemOpen
  \bibfield  {author} {\bibinfo {author} {\bibfnamefont {T.}~\bibnamefont
  {Vogt}}, \bibinfo {author} {\bibfnamefont {S.}~\bibnamefont {Boden}},
  \bibinfo {author} {\bibfnamefont {A.}~\bibnamefont {Andruszkiewicz}},
  \bibinfo {author} {\bibfnamefont {K.}~\bibnamefont {Eckert}}, \bibinfo
  {author} {\bibfnamefont {S.}~\bibnamefont {Eckert}},\ and\ \bibinfo {author}
  {\bibfnamefont {G.}~\bibnamefont {Gerbeth}},\ }\bibfield  {title} {\bibinfo
  {title} {Detection of gas entrainment into liquid metals},\ }\href
  {https://doi.org/10.1016/j.nucengdes.2015.07.072} {\bibfield  {journal}
  {\bibinfo  {journal} {Nucl. Eng. Des.}\ }\textbf {\bibinfo {volume} {294}},\
  \bibinfo {pages} {16} (\bibinfo {year} {2015})}\BibitemShut {NoStop}%
\bibitem [{\citenamefont {Burguete}\ and\ \citenamefont
  {Miranda}(2012)}]{Burguete2012}%
  \BibitemOpen
  \bibfield  {author} {\bibinfo {author} {\bibfnamefont {J.}~\bibnamefont
  {Burguete}}\ and\ \bibinfo {author} {\bibfnamefont {M.}~\bibnamefont
  {Miranda}},\ }\bibfield  {title} {\bibinfo {title} {Instabilities of
  conducting fluid layers in cylindrical cells under the external forcing of
  weak magnetic fields},\ }\href@noop {} {\bibfield  {journal} {\bibinfo
  {journal} {Magnetohydrodynamics}\ }\textbf {\bibinfo {volume} {48}},\
  \bibinfo {pages} {69} (\bibinfo {year} {2012})}\BibitemShut {NoStop}%
\bibitem [{\citenamefont {Grants}(2021)}]{Grants2021a}%
  \BibitemOpen
  \bibfield  {author} {\bibinfo {author} {\bibfnamefont {I.}~\bibnamefont
  {Grants}},\ }\bibfield  {title} {\bibinfo {title} {Rotating magnetic
  dipole-driven flows in a conducting liquid cylinder},\ }\href
  {https://doi.org/10.1063/5.0047240} {\bibfield  {journal} {\bibinfo
  {journal} {Phys. Fluids}\ }\textbf {\bibinfo {volume} {33}},\ \bibinfo
  {pages} {055115} (\bibinfo {year} {2021})}\BibitemShut {NoStop}%
\bibitem [{\citenamefont {Denisov}\ \emph {et~al.}(2010)\citenamefont
  {Denisov}, \citenamefont {Dolgikh}, \citenamefont {Kolesnichenko},
  \citenamefont {Khalilov}, \citenamefont {Khripchenko}, \citenamefont
  {Verhille}, \citenamefont {Plihon},\ and\ \citenamefont
  {Pinton}}]{Denisov2010}%
  \BibitemOpen
  \bibfield  {author} {\bibinfo {author} {\bibfnamefont {S.}~\bibnamefont
  {Denisov}}, \bibinfo {author} {\bibfnamefont {V.}~\bibnamefont {Dolgikh}},
  \bibinfo {author} {\bibfnamefont {I.}~\bibnamefont {Kolesnichenko}}, \bibinfo
  {author} {\bibfnamefont {R.}~\bibnamefont {Khalilov}}, \bibinfo {author}
  {\bibfnamefont {S.}~\bibnamefont {Khripchenko}}, \bibinfo {author}
  {\bibfnamefont {G.}~\bibnamefont {Verhille}}, \bibinfo {author}
  {\bibfnamefont {N.}~\bibnamefont {Plihon}},\ and\ \bibinfo {author}
  {\bibfnamefont {J.-F.}\ \bibnamefont {Pinton}},\ }\bibfield  {title}
  {\bibinfo {title} {Flow of liquid metal in a cylindrical crystallizer
  generating two-directional {{MHD-stirring}}},\ }\href@noop {} {\bibfield
  {journal} {\bibinfo  {journal} {Magnetohydrodynamics}\ }\textbf {\bibinfo
  {volume} {46}},\ \bibinfo {pages} {69} (\bibinfo {year} {2010})}\BibitemShut
  {NoStop}%
\bibitem [{\citenamefont {Horstmann}\ \emph {et~al.}(2019)\citenamefont
  {Horstmann}, \citenamefont {Wylega},\ and\ \citenamefont
  {Weier}}]{Horstmann2019}%
  \BibitemOpen
  \bibfield  {author} {\bibinfo {author} {\bibfnamefont {G.~M.}\ \bibnamefont
  {Horstmann}}, \bibinfo {author} {\bibfnamefont {M.}~\bibnamefont {Wylega}},\
  and\ \bibinfo {author} {\bibfnamefont {T.}~\bibnamefont {Weier}},\ }\bibfield
   {title} {\bibinfo {title} {Measurement of interfacial wave dynamics in
  orbitally shaken cylindrical containers using ultrasound pulse-echo
  techniques},\ }\href {https://doi.org/10.1007/s00348-019-2699-0} {\bibfield
  {journal} {\bibinfo  {journal} {Exp. Fluids}\ }\textbf {\bibinfo {volume}
  {60}},\ \bibinfo {pages} {56} (\bibinfo {year} {2019})}\BibitemShut {NoStop}%
\end{thebibliography}%

\end{document}